\documentclass{rspublic}

\usepackage{epsf}
\usepackage{longtable}
\usepackage{natbib}
\usepackage{epstopdf}
\usepackage{graphicx}

\begin{document}

\title{The Role of Partial Ionization Effects in the Chromosphere}

\author[J. Martinez-Sykora]{Juan Mart\'inez-Sykora $^{1,2}$ \& 
Bart De Pontieu $^{2,3}$ \&
Viggo Hansteen $^{3,2}$ \&
Mats Carlsson $^{3}$}

\affiliation{$^1$ Bay Area Environmental Research Institute, Petaluma, CA \\
$^2$ Lockheed Martin Solar and Astrophysics Laboratory, Palo Alto, CA 94304 \\
$^3$ Institute of Theoretical Astrophysics, University of Oslo, P.O. Box 1029 Blindern, N-0315 Oslo, Norway}

\label{firstpage}

\maketitle

\begin{abstract}{Magnetohydrodynamics MHD ---Methods: numerical  --- Sun: atmosphere --- Sun: magnetic field}

The energy for the coronal heating must be provided from the convection
zone. However, the amount and the method by which this energy is transferred
into the corona depends on the properties of the lower atmosphere and
the corona itself. We review: 1) how the energy 
could be built in the lower solar atmosphere; 2) how this energy is transferred 
through the solar atmosphere; 
 and 3) how the energy is finally dissipated in the chromosphere 
 and/or corona. Any mechanism of energy transport has to deal 
with the various physical processes in the lower atmosphere. We will
focus on a physical process that seems to be highly important 
in the chromosphere and not deeply studied until recently: the ion-neutral
interaction effects in the chromosphere. We review the relevance 
and the role of the partial ionization in the chromosphere
and show that this process actually impacts considerably
the outer solar atmosphere. We include analysis of 
our 2.5D radiative MHD simulations with the {\it Bifrost} code 
(\citealt{Gudiksen:2011qy}) including the partial ionization 
effects on the chromosphere and corona and thermal 
conduction along magnetic field lines. The photosphere, chromosphere 
and transition region are partially ionized and the interaction between ionized particles 
and neutral particles has important consequences on the magneto-thermodynamics 
of these layers. The partial ionization effects are treated 
using generalized Ohm's law, i.e., we consider the Hall term and the 
ambipolar diffusion (Pedersen dissipation) in the induction equation. The 
interaction between the different species affects the modeled atmosphere 
as follows: 1) the ambipolar diffusion dissipates magnetic energy and 
increases the minimum temperature in the chromosphere; 2) the upper 
chromosphere may get heated and expanded over a greater range of heights. 
These processes reveal appreciable differences between the modeled atmospheres of 
simulations with and without ion-neutral interaction effects.
\end{abstract}

\section{Coronal heating vs heating concentrated in the low atmosphere}

One of the recent and heated discussions on chromospheric and corona
heating concerns the location of magnetic energy release in 
the solar atmosphere. Is it highly dependent on height? What is the
role of flux emergence and what is the role of the magnetic field topology?

One dimensional hydrodynamic simulations can cast some light on this
issue. For example, \cite{Testa17102014} use 1D models to prove that nano-flares
situated in the corona can spread energy via non-thermal electron
beams rather than through thermal conduction. 
As another example, \cite{Klimchuk:2014fk}
compared synthetic data obtained from 1D multi parametric hydro simulations with 
the Atmospheric Imaging Assembly observations (AIA, \citealt{Lemen:2012uq,Boerner:2012qf})
on board of the Solar Dynamics Observatory (SDO, \citealt{Pesnell:2012nr}). 
They claimed that their 
synthetic data can not reproduce the observables unless they consider 
quasi-steady coronal heating or coronal nano flares situated well
above the chromosphere. This particular case highlights the limits of
a 1D hydrodynamic approach: advanced 3D simulations of heating in
association with chromospheric jets indicate that the complex 3D
topology needs to be taken into account to understand the impact on
the corona (\citealt{Hansteen:2010uq}).

It is clear that studying the emergence and topology of the
magnetic field and the likely sites of energy release requires full 2D or 3D 
models covering the region from the photosphere to the corona. 
Sophisticated simulations that include thermal conduction along
magnetic field lines (\citealt{Gudiksen+Nordlund2004}), radiative losses in the
 corona and chromosphere (\citealt{Hansteen:2007dt})
are now well established tools and provide quite sophisticated comparisons with 
observations (\citealt{Hansteen:2010uq,Bingert:2011fk}). These models show
that the average heating per particle coming from the released magnetic energy  
is largely concentrated in the region between the upper chromosphere and the lower corona. 
Synthetic observations generated from these models reveal overall
similarities with observed intensities, line shifts, 
widths, and profiles (see e.g., \citealt{Gudiksen:2005lr,
Peter:2006zk,Olluri:2013uq,Hansteen17102014}). 
For instance, \cite{Hansteen:2010uq} use the Bifrost code
(\citealt{Gudiksen:2011qy}) and achieve rather good agreement 
between synthetic and observed average redshifts for transition region spectral lines. They argue that this is 
due to episodic heating being localized in the lower atmosphere (somewhere between the middle 
chromosphere and the lower corona). 

Simulations by \cite{Martinez-Sykora:2011oq} also reveal another observable due to the presence of localized episodic heating:
they studied the asymmetries of the EUV spectral profiles using 
so-called Red-Blue asymmetry (RB) analysis (\citealt{De-Pontieu:2009fk}). 
The largest blue-ward asymmetries are for transition region EUV profiles 
 (\citealt{Kjeldseth-Moe:1977lr,Peter:2001qy,Hara:2008ad,De-Pontieu:2009fk,
McIntosh:2009yf,Peter:2010fk,Tian:2011dq}).  \cite{Martinez-Sykora:2011oq} 
find that the synthetic RB analysis of various EUV lines from their simulations reproduces, 
qualitatively, the observations taken by Solar Ultraviolet 
Measurements of Emitted Radiation
(SUMER, \citealt{Wilhelm:1995fk}) on board of the Solar and Heliospheric Observatory 
(SOHO, \citealt{Domingo:1995sf}) and the
EUV Imaging Spectrometer (EIS, \citealt{Culhane:2000lr}) on board of Hinode 
(\citealt{2007SoPh..243....3K}).
Furthermore, they find that the episodic heating produces
a larger range of velocities within the transition region than in the corona. 
Moreover, the simulations by \cite{Martinez-Sykora:2011oq} include small scale flux emergence, and find that the average volumetric heating is still localized largely within the transition region
(\citealt{Martinez-Sykora:2009rw,Hansteen:2010uq}).
Despite this, they find a quantitative disagreement with observations when looking at
line asymmetries, most likely due to the lack of vigorous dynamics in the simulations. 
A word of caution is in order, these synthetic spectral results and models may 
be significantly impacted by the simplified magnetic field configuration utilized
in the simulations. 
The penalty paid in performing full 3D simulations is the high
resistivity required to keep structures larger than the numerical grid
scale. Small scale plasma physics can therefore not be fully resolved, and the
treatment of phenomena such as the damping of shocks and the magnetic
reconnection process occur at scales much larger than in nature.
Therefore, one must be aware of the limitations of not resolving some of the physical 
processes which may impact the transport of the 
magnetic field stresses (built by the convective motions) to greater
heights in the upper atmosphere.

\section{How to build stress in the atmosphere}~\label{sec:photo}

Let us start at the source: it is well known and broadly accepted that  
convective motions build stresses in the magnetic field in the atmosphere.
In the quiet sun and coronal holes, there are at least five different mechanisms  
in the photosphere that could build stress or transport energy into the upper layers. 
The first is the magneto-acoustic shocks produced by the convective motions, 
p-modes, overshooting, convective collapse, etc  
(\citealt{Carlsson:1992kl,Carlsson:1995ai,Bellot-Rubio:2001ez,
Stein:2006qy,Skartlien:2000lr,Hansteen+DePontieu2006,De-Pontieu:2007cr,
Martinez-Sykora:2009kl,Kato:2011dz} among many others),
or even by reconnection or magnetic energy release in the photosphere or 
lower chromosphere processes (\citealt{Martinez-Sykora:2009kl}). 
The second lies in waves, where the restoring force is ``magnetic''
such as kink-, Aflv\'en-, or other transversal waves, in the following
called ``alfvenic'', produced by reconnection processes and/or convective motions
(\citealt{De-Pontieu:2007bd,Jess:2009bh,Goossens:2009rw,McIntosh:2011fk,van-Ballegooijen:2011fp,De-Moortel:2012xe,De-Pontieu:2014fv}). 
The third is by small scale flux emergence (\citealt{Centeno:2007lr,
Martinez-Gonzalez:2009rp,Martinez-Gonzalez:2012cl,Sainz-Dalda:2012qf} among others). In short,  
these observations reveal that small scale loops occur throughout the
photosphere, e.g.,  \cite{Martinez-Gonzalez:2012cl} observations of the quiet sun 
revealed the emergence of small scale magnetic field loops in the
photosphere at a rate of the order of 
0.2 loops h$^{-1}$ arcsec$^{-2}$. 
The fourth mechanism is through mass flux. The proponents 
note that mass flux also can contribute 
to transport or build energy in the upper layers of the atmosphere, for instance photospheric supersonic 
jets observed by \cite{Bellot-Rubio:2009km} with the Solar Optical Telescope 
(SOT, \citealt{Tsuneta:2008kc}) on board of Hinode and by \cite{Borrero:2012wa} 
with IMAX on board of the balloon-borne solar observatory SUNRISE 
(\citealt{Solanki:2010pt}). Another example of mass flux contributing to
energy supply in the outer layers is that of spicules, with
\cite{De-Pontieu:2011lr} suggesting that some of the spicular mass flux
is heated to coronal temperatures, based on observations with Hinode/SOT and 
SDO/AIA (see also \citealt{Martinez-Sykora:2011uq,Martinez-Sykora:2013ys}).
Finally, the motion of magnetic flux elements in the photosphere,
driven by convective motions will easily build magnetic field stresses
that propagate into the corona, as illustrated with the well known
cartoon by \cite{Parker:1983fj}. 
Note, these mechanisms are not completely independent, e.g., flux emergence 
can also drive waves. Another possible source of the mechanical energy
needed to build extra current is the highly dynamic chromosphere (see below). 

We would like to describe in greater detail two of the sources listed above that we consider 
of great interest in this review.
The first one concerns small scale flux emergence. Small bipole structures in 
the Stokes V magnetograms suggest tiny loops within the granules such as 
observed by \cite{Lites:1996kh,Lites:1998cr}, \cite{De-Pontieu:2002by} and later by 
\cite{Centeno:2007lr}, \cite{Ishikawa:2008cq}, \cite{Martinez-Gonzalez:2009rp}, 
and \cite{Gomory:2010gb}, 
among others. A large number of small scale loops are found to occur in the 
quiet sun. In addition, strong asymmetric Stokes V profiles suggest 
structures of unresolved, even tinier, loops 
(\citealt{Grossmann-Doerth:2000wt,Sainz-Dalda:2012qf,Viticchie:2012ez}). 
\cite{Sainz-Dalda:2012qf} combined inversion with synthetic profiles from 
3D radiative MHD simulations in order to explain the strong asymmetry 
profiles. 
\cite{Viticchie:2011si} categorized a large sample of different 
asymmetries and inverted the corresponding Stokes profiles. 
Many of these strong asymmetries suggest the presence of 
two different polarities of magnetic field within the same pixel. Finally, 
various types of observations reveal pervasive  
large amount of horizontal magnetic flux at the photosphere 
(\citealt{Lites:1996kh,Lites:2008ss,Borrero:2011qc,Bellot-Rubio:2012rq,Stenflo:2013fc} among others). 
\cite{Martinez-Gonzalez:2012cl} suggested that 30\% of these tiny loops are 
able to reach the chromosphere (observations, for instance, by
\citealt{Centeno:2007lr,Martinez-Gonzalez:2009rp,Guglielmino:2010lr,Ortiz:2014wj} also reveal 
the various chromospheric signals of flux emergence). 
 
The second aspect we would like to discuss in a bit more detail is the motion 
of the magnetic flux elements. Magnetic field elements move around
due to the convective motions, collapsing granules and emerging new convective 
cells. The magnetic field elements evolve with the convective motions, merging 
elements of the same polarity, fragmenting magnetic elements, and canceling 
magnetic elements of different polarity (\citealt{Schrijver:1997xu}). It is important to  
understand how the convective motion is braiding and tangling the magnetic 
field in the solar atmosphere (\citealt{Lamb:2008hi,DeForest:2007ak,
Zhou:2010vh,Giannattasio:2014ts} among others) since the 
convective motion could build magnetic energy and thereafter this energy
may dissipate into kinetic and thermal energy. 
These authors studied the dynamic properties of the magnetic field elements
at the photosphere in the quiet sun. Most of the identified magnetic network 
elements seem to form due to unresolved small scale shredding, 
coalescence of previous magnetic flux, magnetic field cancelation and 
emergence.  

Moreover, it is crucial to calculate the magnetic elements that have merged, canceled 
and/or fragmented in order to understand how the magnetic field is stressed in the 
atmosphere (\citealt{Schrijver:1997xu,Iida:2012rf,Zhou:2013bl,Gosic:2014ef}). 
\cite{Gosic:2014ef} using long time series of Hinode/SOT observations 
followed and identified different internetwork elements, and 
noting which of these reach the network. Their conclusions are that internetwork 
elements interact with network elements and thereby modify its net
magnetic flux. 
Only a tiny part of the internetwork elements that interact with the network 
comes from flux emergence and most of the network flux 
comes from shredding and coalescence of previous internetwork magnetic 
flux. The internetwork magnetic flux replaces the network flux in a time period of 9 hours. 
On average, 40\% of the 
internetwork individual elements ends up, eventually, in the network (\citealt{Gosic:2014ef}). 

\section{The various ``filters" of the various atmospheric layers}~\label{sec:filters}

As discussed above, convective motions have several methods of generating enough energy to
heat the corona. The next question becomes:  
What happens to this energy for each source mentioned 
in the previous section as it propagates through
the solar atmosphere? In the solar atmosphere there are many physical processes 
that play different roles in different layers. Likewise, one must also
consider whether the transport of energy can trigger other mechanisms 
that contribute further to increasing the magnetic stresses in the
atmosphere. Finally, there remains the question of where magnetic
energy is released for each source, if it is released at all. 

In short, in order to reach the corona, the energy must go through: 
\begin{itemize}
\item The photosphere, where a large amount of the convective energy is released 
by radiative transport, and the atmosphere is highly
stratified with a pressure scale height of the order of less than 100~km. The upper photosphere is
dominated by large velocities due to convective overshoot.
\item The chromosphere is the interface layer between the 
solar surface (photosphere) and the million degree corona. The chromosphere is of 
great interest because, in principle, it contains enough non-thermal 
energy to heat the entire corona. Many complex physical processes play a role in the 
chromosphere: {\it a)} the plasma is in non-local thermodynamic equilibrium (NLTE), {\it b)} 
the radiation is optically thick, {\it c)} radiation is highly
scattered, {\it d)} ionization is not necessarily  
in equilibrium, {\it e)} the gas is partially ionized, {\it f)} in the upper chromosphere, transition 
region and corona, the thermal conductive flux propagates along the magnetic field lines. 
In addition, the chromosphere is also distinguished by several transitions such as 
from non-magnetized to magnetized, and from partial to full
ionization, from high to low plasma $\beta$ (where plasma 
$\beta=\frac{4\pi P}{B^2}$ is the ratio between the gas pressure and magnetic 
pressure), from optically thick to optically thin, etc.
Note that many of the processes (especially from the items {\it a} to {\it d}) lead to 
highly complex interpretations of imaging and spectral observations 
(e.g., \citealt{Leenaarts:2012cr,Leenaarts:2013ij}). 
\item The transition region, located between the chromosphere and corona, is 
optically thin, magnetically dominated, with steep temperature
gradients, filled with shocks and flows approaching or exceeding the
local speed of sound, and where, as mentioned, thermal conduction 
is a highly important process. In this region we cannot expect ions to
be in ionization equilibrium and for low density regions such as
coronal holes we may also find that ion and electron temperatures
differ (\citealt{Hansteen:1993fy,Lie-Svendsen:2001la}).
\item Finally, in the corona, which generally has low plasma beta, radiation is
optically thin, and, as in the transition region, the 
ionization of some heavy ions is not in equilibrium 
(\citealt{Joselyn:1977th,Hansteen:1993fy,Bradshaw:2006nx,Olluri:2013fu}), 
background radiation heats lower layers, the 
timescales of several important processes are very short and
non-thermal physics may be vital in describing these processes. 
\end{itemize}

Any kind of magnetic stress built by the convective motions 
that reaches the corona will be strongly affected by the various  
processes mentioned above. For instance, the
chromosphere appears to be filled with magnetic field, even in the quiet
sun, despite the fact that the photosphere is sub-adiabatic, which thwarts the 
expansion of emerging magnetic flux into the regions above
(\citealt{Acheson:1979lr,archontis2004}). Therefore, it remains 
unknown under which conditions small-scale flux emerges into the
chromosphere.   The challenge facing the researcher is thus to combine all
these ingredients into a model of the entire solar atmosphere, taking
into account all important effects while discarding those that have
little impact on the buildup, transport and dissipation of the energy
flux heating the outer atmosphere.

\section{Partial ionization effects}

In this paper we will concentrate on ion-neutral interaction effects
among all the possible physical processes from the photosphere 
to the lower corona. The chromosphere is partially ionized: 
ions are coupled to the magnetic field, whereas neutrals are not directly 
affected by the presence of the magnetic field and they can move ``freely". Sufficient 
collisions between ions and neutrals will couple the neutrals to the magnetic field, 
while at the same time to some extent allowing ions to slip across the
field. As a result, the magnetic field can diffuse and magnetic energy
will be dissipated into thermal energy. In order to model this
scenario, the fluid should be treated as consisting of two or three fluids, i.e., 
ion, neutral and/or electron particles. To solve the equations for all
three species is computationally 
highly demanding. Therefore two or three fluid codes focus on specific and localized problems 
(\citealt{Smith:2008oa,Sakai:2009le,Meier:2011xd,Leake:2012pr} among others)
instead of modeling large portions of the solar atmosphere, i.e.,
instead of including the convection zone, photosphere,
chromosphere, transition region and corona. Additionally these codes
rarely consider many of the other important physical processes in
the solar atmosphere such as radiative transfer,  
thermal conduction along the magnetic field lines, etc, which 
is the matter of interest we wish to cover in this review (see previous sections).

\subsection{The Hall Term and Ambipolar Diffusion}

Fortunately, the chromosphere seems to be sufficiently collisional to
allow us to convert the three fluid problem into a single MHD fluid
while still taking into account the most important partial ionization effects. 
In order to do this one assumes that 1) collisions are sufficiently 
numerous and the ion and electron temperatures are the same,  2) the
relevant resolvable timescales one wants are assumed to be greater 
than the collision times between ions and neutrals 
(\citealt{cowling1957,Braginskii:1965ul,Parker:2007lr,Pandey:2008qy,Leake:2013fk}). 
According to the comparison of the collision rates with timescales of various relevant
physical processes from the self-consistent 2D radiative-MHD atmospheric models done 
by \cite{Martinez-Sykora:2012uq} using the Bifrost code (\citealt{Gudiksen:2011qy}), 
the approximations mentioned above seem to be fulfilled 
in the chromosphere most of the time. 

In this case, the induction equation expands into the so-called generalized 
induction equation which includes at least two new terms, i.e., the Hall term and the 
ambipolar diffusion (or Pedersen dissipation) term \citep{cowling1957,Braginskii:1965ul} 
as follows:

\begin{eqnarray}
\frac{\partial {\bf B}}{\partial t} = && \nabla \times \left[{\bf u \times B} -  \eta_{ohm} {\bf J}
 - \frac{\eta_{hall}}{ |B|} {\bf J \times B} + \frac{\eta_{amb}}{ B^2} ({\bf J \times B}) 
 \times {\bf B} \right]\label{eq:faradtot2}
\end{eqnarray}

\noindent
where ${\bf B}$, ${\bf J}$, ${\bf u}$, and $\eta_{ohm}$ are magnetic field, current density, 
velocity field, and the ohmic diffusion, respectively. The new ``diffusion'' terms are: 

\begin{eqnarray}
\eta_{ohm} & = & \frac{ m_e (\nu_{ei}+\nu_{en})}{q_e^2 n_e}\\
\eta_{hall} & = & \frac{ |B|}{q_e n_e}\\
\eta_{amb} & = & \frac{(|B| \rho_n/\rho)^2}{\rho_i \nu_{in}} = \frac{(|B|
  \rho_n/\rho)^2}{\rho_n \nu_{ni}}
\end{eqnarray}

\noindent
where $\rho_i$, $\rho_n$, $\rho$, $\nu_{in}$, $\nu_{ni}$, $n_{e}$, and $q_e$
are ion density, neutral density, total density, ion-neutral collision frequency, 
neutral-ion collision frequency, electron number density and electron charge, 
respectively. Note that $\eta_{hall}$ is not a diffusive or dissipative term in contrast 
to $\eta_{ohm}$ and $\eta_{amb}$. The ambipolar diffusion should not be confused with 
the so-called ambipolar drift used in plasma physics. For plasma physics, ambipolar 
drift or electric field is due to the relative 
velocity or temperature difference between ions and electrons, e.g., 
\cite{Arefiev:2008vn}. Ambipolar diffusion is also 
referred to as Pedersen dissipation.  

In short, the ohmic diffusion can change directly the magnetic field topology, i.e., change the connectivity
of the field lines with reconnection, but the Hall and the ambipolar terms cannot, without the help of 
ohmic diffusion, since the former term does not follow the formalism of 
${\bf u} \times {\bf B}$ (\citealt{Biskamp:2005uq}) and the other two follow it, where 
${\bf u}$ for the Hall term is ${\bf J}$
and for the ambipolar term is ${\bf J}\times {\bf B}$. The ohmic (first term in the 
right hand side of the following expression) and ambipolar diffusion (second 
term in the right hand side of the following expression) can dissipate magnetic 
energy into thermal energy as follows:

\begin{eqnarray}
\frac{\partial e}{\partial t} \propto \eta_{ohm} J^2 + \eta_{amb} \frac{|{\bf J}\times {\bf B}|^2}{|B|^2}
\end{eqnarray}

\noindent (see \citealt{Parker:2007lr} for details). 

The physical process behind the Hall term is that in a fully ionized plasma 
as in the corona, the relative immobility of ions relative to electrons in 
response to the Lorentz force leads to an electric field parallel to ${\bf J}\times {\bf B}$. 
The Hall term for weakly ionized plasma is created due to collisions between ions 
and neutrals. The neutrals decouple ions from the magnetic field lines leading again 
to an electric field parallel to ${\bf J}\times {\bf B}$. 
The physical process behind the ambipolar diffusion is that neutrals do not 
experience the Lorentz force. As a result, neutrals decouple 
from the magnetic field and allow the magnetic field to diffuse through the neutral 
gas (\citealt{Braginskii:1965ul}).

\subsection{Impact of the Ion-Neutral Interaction Effects on Simplified
Physical Processes}

Partial ionization effects impact various physical processes that play a role
in the chromosphere and prominences. For instance, the partial 
ionization effects in the chromosphere are known to allow ion-neutral 
collisions to dissipate alfvenic waves (\citealt{de-Pontieu:1998lr,de-Pontieu:1999kx}). 
However, the picture is far from complete, since it is unknown if this dissipation 
plays a major role in heating the chromosphere 
(\citealt{Biermann:1948ye,Leake:2005rt,Hasan:2008ys,Song:2011ly,Madsen:2014ly}). 
Most of these studies obtain different high frequency cut-offs for alfvenic waves for different 
features (spicules, filaments, prominences, thin flux tubes, etc) using fixed values for
the ion-neutral collision frequency of semi-empirical models such as the VAL-C model 
(\citealt{De-Pontieu:2001fj,Leake:2005rt,Forteza:2007zp}). This dissipation contributes to the 
thermal and dynamic energy budget. In fact, as mentioned in the previous section, 
electrical currents perpendicular to the magnetic field can be dissipated by 
ambipolar diffusion and convert magnetic energy into thermal energy 
(\citealt{Arber:2009ve,Goodman:2011wq,Khomenko:2012bh,Goodman:2012vn}). 
\cite{Soler:2009hw,Soler:2012hb} deduced that 
kink waves are damped by diffusion parallel to the magnetic field and Alfv\'en waves 
by perpendicular diffusion. For short wavelengths, kink waves are converted
into Alfv\'en waves due to the 
collisions between ion and neutrals. Ohmic dissipation dominates
for long wavelengths. In fact, all the diffusive mechanisms must be considered, 
and missing any of the inertial, Hall and ambipolar terms will give incorrect results 
(\citealt{Khodachenko:2004vn,Khodachenko:2006kx}). 
\citealt{Zaqarashvili:2012iw} deduced that there is no cut-off frequency 
for Alfv\'en waves, only a damping mechanism due to the ion-neutral interaction effects, when 
one is taking into account both the inertial and Hall term (where the inertial term is defined as the time variation of the 
relative velocity perturbations perpendicular to the unperturbed magnetic field between 
ions and neutrals). However, 
according to \citet{Vranjes:2014fk}, this result might be due to the assumption of neglecting the 
exchange of momentum between ions and neutrals. However, one must be really careful with all these
results since the ion-neutral collision frequency has been calculated in a very simplified 
manner and a proper treatment of the collision between ion and neutrals may provide 
different results (see below). In fact, even in semi-empirical 
models, using different methods to calculate the ion-neutral collision frequency provide
very different values for the damping and ambipolar diffusion terms (\citealt{De-Pontieu:2001fj}). 

Ambipolar diffusion allows magnetic field to diffuse into the atmosphere 
(\citealt{Leake2006,Arber:2007yf,Leake:2013dq}). Since neutrals move through the 
expanding magnetic field in the chromosphere, the magnetic field lines lift less 
matter. As a result the ambipolar diffusion inhibits the 
Rayleigh-Taylor (RT) instability. In contrast, a completely different 
configuration, such as the 2.5D simulations of prominences 
including partial ionization, shows an increase of small scale velocities as a result of
the non-linearity of the RT instability, in contrast to single fluid 
simulations (\citealt{Diaz:2014dq,Khomenko:2014zr}). Note that, 
 the ambipolar diffusion can have effects of opposite nature: it can either inhibit the RT instability 
or increase the small scale structures of the RT instability depending on the spatial distribution 
of ambipolar diffusion, i.e., compare \cite{Arber:2007yf} with \cite{Diaz:2014dq} and 
\cite{Khomenko:2014zr}. Since the 
ambipolar diffusion is highly dependent on the thermal properties of the 
plasma, the ion-neutral collision frequency needs to be calculated directly from the 
thermal properties of the plasma and the energy balance must be solved self-consistently. 

Regardless of the assumptions of the background atmosphere and the
calculation of the ion-neutral collision frequency, it is clear that 
partial ionization effects play an important role in these processes. 

As mentioned before, ion-neutral interaction effects cannot change the topology 
of the magnetic field, but they can impact the reconnection rate. 
On one hand, ambipolar diffusion causes sharp structures to evolve around the
reconnection region, thus setting the scene for tearing
reconnection, accelerating the reconnection 
rate or, depending on the configuration of the magnetic 
field and the diffusivities, singularities that initiate
reconnection. On the other hand, 
electron pressure and ohmic dissipation act against the formation of sharp currents, 
i.e., the reconnection rate depends on the ionization rate and ion-neutral collision frequency
(\citealt{Brandenburg:1994qy,Brandenburg:1995jb}). When the ionization decreases, 
the reconnection rate decreases too (\citealt{Smith:2008oa,Sakai:2009le}). Note that  
the two fluid simulations from the latter two references used fixed ionization fractions 
and did not take into account the ionization/recombination rates. In addition, they did 
not clarify if their parametric study is for a fixed density or a fixed number of ions since 
an increase of density can lead to a decrease of the reconnection rate. 
In a reconnection process in a weakly ionized plasma when ions and neutrals are decoupled, 
it is found that an excess of ions can accumulate in the reconnection region, as shown with 
the simulations done by \cite{Vishniac:1999zk} and \cite{Lazarian:2004xi}. 
The recombination in the reconnection region removes ions which
combined with the alfvenic flows can lead to fast reconnection. 
This has been tested with 2D two fluid
simulations taking into account the ionization/recombination, optically thin radiative transfer and
collisional heating by \cite{Leake:2012pr,Leake:2013qz}. 
An important remark from the latter authors is that it 
is crucial to treat the ambipolar diffusion while at the same time
solving the energy balance including the ionization/recombination since it can change 
drastically the impact of the ambipolar diffusion on the reconnection process. 
While the results described above indicate that ion-neutral
effects are an important ingredient in a variety of processes in the
solar atmosphere, the current simplified approaches to the energy
balance significantly limit the applicability of these results in the 
solar atmosphere. 

All the studies mentioned in this section are focussed on specific driving processes. 
These studies are extremely useful to discern the physical mechanism in each specific 
problem. However, in order to place these processes in the context of the
solar atmosphere, we need to combine this knowledge with self-consistent radiative MHD models that have
the proper convective drivers (see Section~\ref{sec:photo}) and that
include the relevant physical processes in each layer of the atmosphere, as done by 
\cite{Hansteen:2007dt}, \cite{abbett2007}, \cite{paper1}, and \cite{Hansteen:2010uq}, 
amongst others. The missing ingredient in these models has so far been
ion-neutral interactions which should be calculated self-consistently
with the thermal properties (such as done by \citealt{Leake:2012pr,Leake:2013qz} 
for simplified magnetic reconnection simulations).
The Hall effect in the photosphere has been implemented following 
this approach by \cite{Cheung:2012uq}. The preliminary work done in 
\cite{Martinez-Sykora:2012uq} added ambipolar diffusion to the
description of the outer solar atmosphere. We describe this type of simulations 
including ambipolar diffusion in the next sections. 

\subsection{Importance of the Ambipolar Diffusion in the Chromosphere}

In the previous section we described several aspects of how ion-neutral 
interaction impacts various  
physical processes in the chromosphere and prominences. Let us now discuss 
the range of values for the various ``diffusion" terms of the expanded 
induction equation in the lower solar atmosphere. Figure~1 in 
\cite{Khomenko:2012bh} shows, using the semi-empirical VAL-C model and thin 
magnetic flux tube approximation (\citealt{spruit1981,Roberts1978}), the values of 
the various ``diffusion" terms through the atmosphere.
In short, ohmic diffusion dominates below the photosphere. In the photosphere, 
the Hall term is the largest and reaches  $7 \times 10^5$ m$^2$~s$^{-1}$. In the 
chromosphere the largest, by roughly 2 orders of magnitude, is the ambipolar diffusion
($10^8$ m$^2$~s$^{-1}$). 
However, these values become extremely different if 
we consider a radiative MHD simulation such as the one used by, for instance,  
\cite{Gudiksen:2005fk}, \cite{Hansteen:2010uq} (see below). 

In order to calculate the ambipolar diffusion terms, the ion-neutral 
collision frequency must be obtained from the ionization state and 
thermal properties of the plasma. 
For this there are several methods (\citealt{Osterbrock:1961fk, 
von-Steiger:1989uq,Fontenla:1993fj} among others). \cite{Osterbrock:1961fk} uses elastic 
scattering cross sections, \cite{von-Steiger:1989uq} considers hard spheres, 
and \cite{Fontenla:1993fj} uses different sets of elastic collision cross 
sections than  \cite{Osterbrock:1961fk} to determine the ion-neutral collision 
frequency. The final results of these methods differ in the ion-neutral 
collision frequency by an order of magnitude for semi-empirical 
models (\cite{De-Pontieu:2001fj} among others). In a more realistic approach, 
\cite{Martinez-Sykora:2012uq} show that the various methods used to calculate the 
ion-neutral collision frequency differ most in chromospheric regions where the 
ambipolar diffusion is largest. 
A more recent analysis of the ion-neutral collision frequency is done by 
\cite{Vranjes:2013ve} where they include accurate cross sections which vary 
as a function of temperature, due to quantum effects, and between descriptions
of elastic scattering, momentum transfer and viscosity. As mentioned by these authors, 
this work can be improved including the effects of inelastic collisions. 

\begin{figure}[ht]
\centering
 \includegraphics[width=0.99\textwidth]{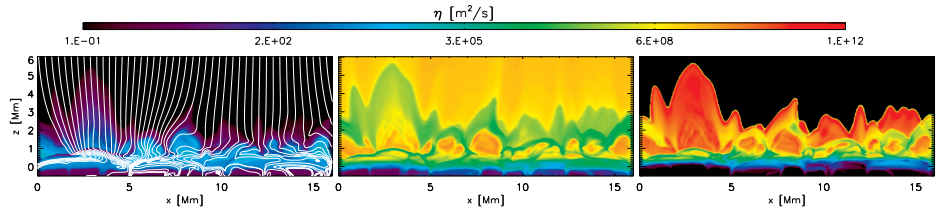}
 \caption{\label{fig:diff} Maps of ohmic diffusion (left), Hall term (middle) and 
 ambipolar diffusion (right) from a snapshot of a 2D radiative-MHD simulation 
 that includes partial ionization effects. The color scheme is on a logarithmic scale. The 
 magnetic field is drawn with white lines in the left panel.}
\end{figure}

Figure~\ref{fig:diff} shows the ohmic diffusion, Hall term and ambipolar 
diffusion where \cite{Osterbrock:1961fk} was used to calculate the 
ion-neutral collision frequency. The simulation is 2.5D and ranges from the upper 
convection zone to the lower corona including radiative transfer with 
scattering from the photosphere to the corona (\citealt{Hayek:2010ac,Carlsson:2012uq}), 
thermal conduction along the magnetic field lines and partial ionization effects 
using the Bifrost code (see \citealt{Martinez-Sykora:2012uq} and references cited within). 
The initial magnetic field is unipolar with a mean unsigned field strength of 
$5$~G in the corona. The convective motions build enough magnetic
field stress to self-consistently maintain the hot corona 
(\citealt{Galsgaard:1996lf,Gudiksen+Nordlund2004,Gudiksen:2005lr,Hansteen:2010uq,Martinez-Sykora:2011oq}). Note that the magnetic 
field is vertical in the corona (left panel), and despite the fact that our 
simulations do not include imposed flux emergence, the dynamics 
in the simulation and the convective motions accumulate horizontal 
magnetic field in the sub-adiabatic photosphere. 
As mentioned above, this horizontal 
field will in principle be constrained to the sub-adiabatic photosphere 
(\citealt{Acheson:1979lr,archontis2004}). In addition, magnetic field 
footpoints will move around. Note that these simulations naturally reproduce several  
of the energy generating drivers discussed in Section~\ref{sec:photo}. Moreover, the energy 
flux through the atmosphere will be modified by the physical processes
considered in the model as it propagates upwards towards the corona as mentioned 
above (see \citealt{Martinez-Sykora:2012uq} and compare with Section~\ref{sec:filters}). 

The ambipolar diffusion is the largest term in the induction equation in the chromosphere by 
almost ten orders of magnitude compared to the ohmic dissipation and 
something between two and six orders of magnitude larger than the Hall term 
(see Figure~\ref{fig:diff} and \citealt{Martinez-Sykora:2012uq}, the latter compares
these terms with the hyper-diffusion intrinsic in the code too.). In
addition, the various terms show large spatial variability. For instance, the ambipolar diffusion changes
at very small scales (within a few hundred km) by up to seven orders of magnitude, e.g., at 
height $z=1$~Mm. This large variability is due to the large changes in the 
ion-neutral collision frequency and in the neutral density when the ionization state 
and thermal properties of the fluid are taken into account. Note that this model is 
self-consistent and the ionization state and thermal properties are set by the various
drivers, such as shocks, and convective motions of the footpoints that build 
magnetic field stress (Section~\ref{sec:photo}) and the physical 
processes implemented in the simulation: radiation, thermal 
conduction along the magnetic field lines, scattering, partial ionization 
effects in the equation of state and ion-neutral interaction effects. 
One important process that has yet to be taken into account in these
models which will change the ohmic diffusion, Hall term, and ambipolar diffusion 
is the effect of time dependent non-equilibrium ionization of hydrogen and helium 
(\citealt{Leenaarts:2007sf,Golding:2014fk}) which are know to be important
in the upper chromosphere. In our models we assume time dependent equilibrium 
ionization. These effects must 
be included in order to draw firm conclusions on the importance of
ion-neutral interactions in this part of the solar atmosphere.

From Figure~\ref{fig:diff}, it is clear that any magnetic energy flux (driven 
by photospheric or chromospheric processes) that is transferred through the 
chromosphere may be strongly influenced and/or diffused by the ambipolar 
diffusion. We can therefore already conclude that the chromosphere may 
diffuse more magnetic field stress than any other atmospheric layer. Note that
the ohmic diffusion is greater in the photosphere and chromosphere than in the 
corona. 

\subsection{Impact on the Solar Atmosphere}

Here we will not describe in detail how ambipolar diffusion affects the various 
processes such as flux emergence and wave propagation in 2D radiative 
MHD simulations. This is outside of the scope of this review and will be 
described in a series of follow up papers. However, we will show the impact 
of ambipolar diffusion on the thermal properties of the solar atmosphere. For 
this we compare different 2D radiative-MHD simulations listed in table~\ref{tab:runs}.

\begin{table}
\centering
 \caption{\label{tab:runs} Description of the simulations}
\longcaption{The left column lists the names of the various 2D simulations, 
and the right column gives a short description of each simulation.}
 \begin{tabular}{|l|l|}
  \hline
\bf{Name} & \bf{Description} \\ \hline \hline
NGOL & without ion-neutral interaction effects \\ \hline
GOL-OS &  with ion-neutral interaction effects using \cite{Osterbrock:1961fk} for $\nu_{in}$ \\ \hline
 GOL-F & with ion-neutral interaction effects using \cite{Fontenla:1993fj} for $\nu_{in}$\\ 
\hline
\end{tabular}
\end{table}

\begin{figure}[ht]
\centering
  \includegraphics[width=0.99\textwidth]{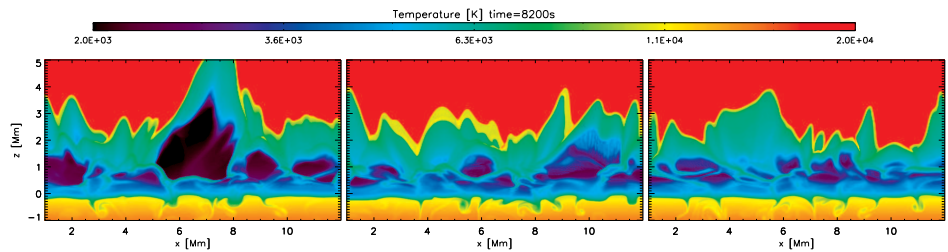} 
 \caption{\label{fig:2dtemp} Temperature maps for the NGOL simulation (left), 
 GOL-OS (middle) and GOL-F (right) reveal differences in the thermal 
 properties. The temperature is shown in logarithmic scale.}
\end{figure}

The different models, i.e., NGOL, GOL-OS and GOL-F, show important 
differences. This means that even with the large artificial diffusion
intrinsic in these codes, the resulting models can discern (at least) some of the effects
produced by the partial ionization interaction effects.
A first look at the 2D temperature maps in Figure~\ref{fig:2dtemp} reveals at 
least two thermal properties that differ between the various simulations: 1) the 
cold chromospheric expanding bubbles have higher temperatures in the GOL 
simulations, and they do not rise as high as in simulations without partial ionization 
effects. 2) The  transition region shows a smaller temperature
gradient (i.e., is spread out over a larger height range) in the 
GOL-OS simulation than in the other two simulations. 

\begin{figure}[ht]
\centering
  \includegraphics[width=0.96\textwidth]{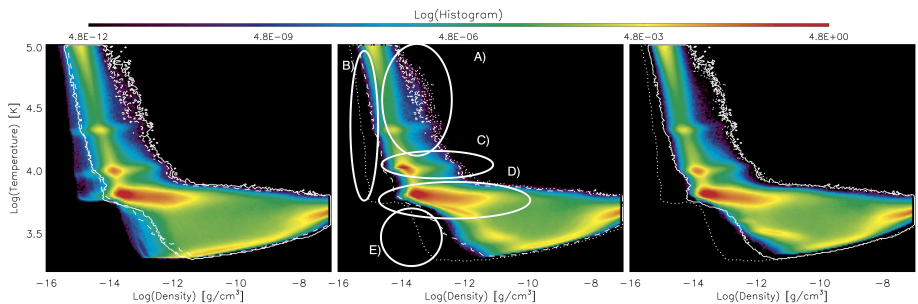} 
 \caption{\label{fig:histtgr} Joint Probability Density Function (JPDF) of temperature 
 (vertical axis)  and density (horizontal axis) integrated over 30 min for simulations 
 NGOL (left), and GOL-OS (middle) and GOL-F (right). The white contours 
 correspond to  the temperature and density regime of the simulation NGOL in 
 dotted line, GOL-OS in solid line and GOL-F in dashed line in order to simplify the
 comparison between the simulations. The circles in the middle panel enhance the 
 differences between the various simulations.}
\end{figure}

Cold chromospheric bubbles are formed by expanding shock fronts that pass 
through the chromosphere. These bubbles contain the lowest temperatures in 
the atmospheric model (\citealt{Leenaarts:2011qy}). Statistically, the simulations 
with partial ionization effects have hotter and denser cold chromospheric expanding 
bubbles than the NGOL simulation. This is shown in the Joint Probability Distribution 
Functions (JPDF) of the density and temperature integrated over 30 minutes in 
Figure~\ref{fig:histtgr} for simulations NGOL (left panel), GOL-OS (middle panel), 
and GOL-F (right panel). Everything below 4000~K 
($\log T < 3.6$) corresponds to the cold chromospheric bubbles. The minimum  
temperature achieved is lower in the NGOL simulation than in the GOL simulations 
(enhanced with white circle E in the middle panel). 
In fact, the coolest temperatures for the NGOL simulation are controlled by an 
{\em ad-hoc} heating term in the Bifrost code which is introduced to avoid reaching
temperatures lower than $\sim 1600$~K since below this threshold the 
tabulated equation of state is no longer accurate (\citealt{Carlsson:2012uq}). 
While this  {\em ad-hoc} heating is crucial for the NGOL 
simulation, it is rarely necessary in both simulations with ion-neutral interaction 
effects since the cold chromospheric 
bubbles are not cold enough to reach the temperature at which the 
{\em ad-hoc} heating starts. The cold chromospheric bubbles in the 
simulations with partial ionization effects are a few 
hundred of degrees hotter than in the NGOL simulation. 
\cite{Leenaarts:2011qy} found no mechanism to prevent extremely low
temperatures from occurring in expanding bubbles within the
standard, single-fluid MHD formulation. 
The joule heating contribution coming from ambipolar diffusion
provides such a mechanism. Observations can help determine the amount
and temperature of such cold plasma in the solar chromosphere
(\citealt{Ayres:1996pi,Penn:2011fk}). The details of the physical processes in 
the chromospheric cold bubbles will be described in a follow up paper. 

The upper chromosphere also shows considerable differences between the three
simulations. One can appreciate that in the GOL simulations there are
a greater number (compared to NGOL) of grid 
points concentrated around the temperature where hydrogen ionizes in the simulations 
(white circle D in the middle
panel of Figure~\ref{fig:histtgr}). Moreover, the 
number of grid points around the temperature where helium ionizes 
($T\sim 10^4$~K) is greater for GOL-OS compared to the other simulations 
(circle C). In fact, GOL-OS shows several grid points with rather 
dense material in the transition region (circle A) while the
transition region is the least dense in NGOL (circle B). GOL-F has the narrowest range of 
densities in the transition region. These discrepancies in the GOL simulations are due to 
the differences in the distribution of 
ambipolar diffusion as a result of the different methods to calculate the ion-neutral collision 
frequency and cross section values. hydrogen and helium time dependent ionization
may spread the amount of plasma over a larger range of temperatures than for 
the case of ionization equilibrium. This latter case leads to the temperature of the plasma 
being localized within the values where hydrogen (circle D) and helium are ionized (circle C).

There are several physical processes that lead to these differences in the upper 
chromosphere. These processes will be described in detail in 
a follow up paper. In short, the expanding cold bubbles, which have large ambipolar 
diffusion, are heated and, it is through these bubbles that the photospheric horizontal magnetic field 
is diffused into the chromosphere. 
From there, the field expands into the corona pushing chromospheric material to higher 
layers. A large percentage of the magnetic stresses due to this expansion of the magnetic field, 
waves and footpoint motions are dissipated in the cold bubble. Moreover, for GOL-OS, 
the diffused field lines coming from the photosphere interact and dissipate magnetic 
energy in the upper chromosphere into thermal energy driving
temperatures up to $\sim 10^{4}$~K. The various  
regions of the atmosphere will impact how 
mechanical and magnetic energy is transported to greater heights in different 
manners, since different processes dominate in the various regions. 
The partial ionization interaction effects also produce a slightly more 
dynamic upper-chromosphere, transition region and 
lower corona as a result of the expansion of the diffused magnetic field and
the interaction of this expanding magnetic field lines with the ambient magnetic field. 
These simulations need to be recalculated taking into account a better ion-neutral collision 
frequency calculation (\citealt{Vranjes:2013ve}).  Including the effects of partial ionization 
is important to many aspects of the physics of the chromosphere and corona, in particular 
those related to the evolution of the magnetic field and the energization of the outer 
atmosphere's plasma.

\section{Summary}

Including the effects of partial ionization is important to many aspects of the physics of the chromosphere and corona, in particular those related to the evolution of the magnetic field and the energization of the outer atmosphereÕs plasma, as the interaction between ionized particles and neutral particles has important consequences on the magneto-thermodynamics. In particular, our simplified 2D radiative MHD models reveal that the ambipolar diffusion dissipates magnetic energy and increases the minimum temperature in the chromosphere and the upper chromosphere may get heated and expanded over a greater range of heights.  

\section{Acknowledgments}

We gratefully acknowledge support by NASA grants NNX11AN98G,
NNM12AB40P and NASA contracts NNM07AA01C (Hinode), and NNG09FA40C 
(IRIS). This research was supported by the Research Council of Norway and by the 
European Research Council under the European Union's Seventh Framework 
Programme (FP7/2007-2013) / ERC Grant agreement nr. 291058.
The simulations have been run on clusters from the Notur project, 
and the Pleiades cluster through the computing project s1061 from the High 
End Computing (HEC) division of NASA. We thankfully acknowledge the 
computer and supercomputer resources of the Research Council of Norway 
through grant 170935/V30 and through grants of computing time from the 
Programme for Supercomputing. This work has benefited from discussions at 
the International Space Science Institute (ISSI) meetings on 
``Heating of the magnetized chromosphere'', on ``Sub-arcsecond Observations 
and Interpretation of the Solar Chromosphere" and ``Coronal Heating - Using 
Observables (flows and emission measure) to Settle the Question of Steady vs. 
Impulsive Heating" where many aspects of this paper were discussed with other 
colleagues.

\bibliographystyle{rspublicnat}
\bibliography{collectionbib2}

\begin{thebibliography}{145}
\providecommand{\natexlab}[1]{#1}
\expandafter\ifx\csname urlstyle\endcsname\relax
  \providecommand{\doi}[1]{doi:\discretionary{}{}{}#1}\else
  \providecommand{\doi}{doi:\discretionary{}{}{}\begingroup
  \urlstyle{rm}\Url}\fi

\bibitem[{{Abbett}(2007)}]{abbett2007}
{Abbett}, W.~P. 2007 {The Magnetic Connection between the Convection Zone and
  Corona in the Quiet Sun}.
\newblock \emph{ApJ}, \textbf{665}, 1469--1488.
\newblock (\doi{10.1086/519788})

\bibitem[{{Acheson}(1979)}]{Acheson:1979lr}
{Acheson}, D.~J. 1979 {Instability by magnetic buoyancy}.
\newblock \emph{SolPhys}, \textbf{62}, 23--50.

\bibitem[{{Arber} \emph{et~al.}(2009){Arber}, {Botha} \&
  {Brady}}]{Arber:2009ve}
{Arber}, T.~D., {Botha}, G.~J.~J. \& {Brady}, C.~S. 2009 {Effect of Solar
  Chromospheric Neutrals on Equilibrium Field Structures}.
\newblock \emph{ApJ}, \textbf{705}, 1183--1188.
\newblock (\doi{10.1088/0004-637X/705/2/1183})

\bibitem[{{Arber} \emph{et~al.}(2007){Arber}, {Haynes} \&
  {Leake}}]{Arber:2007yf}
{Arber}, T.~D., {Haynes}, M. \& {Leake}, J.~E. 2007 {Emergence of a Flux Tube
  through a Partially Ionized Solar Atmosphere}.
\newblock \emph{ApJ}, \textbf{666}, 541--546.
\newblock (\doi{10.1086/520046})

\bibitem[{{Archontis} \emph{et~al.}(2004){Archontis}, {Moreno-Insertis},
  {Galsgaard}, {Hood} \& {O'Shea}}]{archontis2004}
{Archontis}, A., {Moreno-Insertis}, F., {Galsgaard}, K., {Hood}, A. \&
  {O'Shea}, E. 2004 Emergence of magnetic flux from the convection zone into
  the corona.
\newblock \emph{A\&A}, \textbf{426}, 1047--1063.

\bibitem[{{Arefiev} \& {Breizman}(2008)}]{Arefiev:2008vn}
{Arefiev}, A.~V. \& {Breizman}, B.~N. 2008 {Ambipolar acceleration of ions in a
  magnetic nozzle}.
\newblock \emph{Physics of Plasmas}, \textbf{15}(4), 042109.
\newblock (\doi{10.1063/1.2907786})

\bibitem[{{Ayres} \& {Rabin}(1996)}]{Ayres:1996pi}
{Ayres}, T.~R. \& {Rabin}, D. 1996 {Observations of Solar Carbon Monoxide with
  an Imaging Infrared Spectrograph. I. Thermal Bifurcation Revisited}.
\newblock \emph{ApJ}, \textbf{460}, 1042.
\newblock (\doi{10.1086/177031})

\bibitem[{{Bellot Rubio}(2009)}]{Bellot-Rubio:2009km}
{Bellot Rubio}, L.~R. 2009 {Detection of Supersonic Horizontal Flows in the
  Solar Granulation}.
\newblock \emph{ApJ}, \textbf{700}, 284--291.
\newblock (\doi{10.1088/0004-637X/700/1/284})

\bibitem[{{Bellot Rubio} \& {Orozco Su{\'a}rez}(2012)}]{Bellot-Rubio:2012rq}
{Bellot Rubio}, L.~R. \& {Orozco Su{\'a}rez}, D. 2012 {Pervasive Linear
  Polarization Signals in the Quiet Sun}.
\newblock \emph{ApJ}, \textbf{757}, 19.
\newblock (\doi{10.1088/0004-637X/757/1/19})

\bibitem[{{Bellot Rubio} \emph{et~al.}(2001){Bellot Rubio}, {Rodr{\'{\i}}guez
  Hidalgo}, {Collados}, {Khomenko} \& {Ruiz Cobo}}]{Bellot-Rubio:2001ez}
{Bellot Rubio}, L.~R., {Rodr{\'{\i}}guez Hidalgo}, I., {Collados}, M.,
  {Khomenko}, E. \& {Ruiz Cobo}, B. 2001 {Observation of Convective Collapse
  and Upward-moving Shocks in the Quiet Sun}.
\newblock \emph{ApJ}, \textbf{560}, 1010--1019.
\newblock (\doi{10.1086/323063})

\bibitem[{{Biermann}(1948)}]{Biermann:1948ye}
{Biermann}, L. 1948 {{\"U}ber die Ursache der chromosph{\"a}rischen Turbulenz
  und des UV-Exzesses der Sonnenstrahlung}.
\newblock \emph{Zeitschrift fur Astrophysik}, \textbf{25}, 161--+.

\bibitem[{{Bingert} \& {Peter}(2011)}]{Bingert:2011fk}
{Bingert}, S. \& {Peter}, H. 2011 {Intermittent heating in the solar corona
  employing a 3D MHD model}.
\newblock \emph{A\&A}, \textbf{530}, A112.
\newblock (\doi{10.1051/0004-6361/201016019})

\bibitem[{Biskamp(2005)}]{Biskamp:2005uq}
Biskamp, D. 2005 \emph{Magnetic reconnection in plasmas}, vol.~3.
\newblock Cambridge University Press.

\bibitem[{{Boerner} \emph{et~al.}(2012){Boerner}, {Edwards}, {Lemen}, {Rausch},
  {Schrijver}, {Shine}, {Shing}, {Stern}, {Tarbell}
  \emph{et~al.}}]{Boerner:2012qf}
{Boerner}, P., {Edwards}, C., {Lemen}, J., {Rausch}, A., {Schrijver}, C.,
  {Shine}, R., {Shing}, L., {Stern}, R., {Tarbell}, T. \emph{et~al.} 2012
  {Initial Calibration of the Atmospheric Imaging Assembly (AIA) on the Solar
  Dynamics Observatory (SDO)}.
\newblock \emph{SolPhys}, \textbf{275}, 41--66.
\newblock (\doi{10.1007/s11207-011-9804-8})

\bibitem[{{Borrero} \& {Kobel}(2011)}]{Borrero:2011qc}
{Borrero}, J.~M. \& {Kobel}, P. 2011 {Inferring the magnetic field vector in
  the quiet Sun. I. Photon noise and selection criteria}.
\newblock \emph{A\&A}, \textbf{527}, A29.
\newblock (\doi{10.1051/0004-6361/201015634})

\bibitem[{{Borrero} \emph{et~al.}(2012){Borrero}, {Pillet}, {Schlichenmaier},
  {Schmidt}, {Berkefeld}, {Solanki}, {Bonet}, {Iniesta}, {Domingo}
  \emph{et~al.}}]{Borrero:2012wa}
{Borrero}, J.~M., {Pillet}, V.~M., {Schlichenmaier}, R., {Schmidt}, W.,
  {Berkefeld}, T., {Solanki}, S.~K., {Bonet}, J.~A., {Iniesta}, J.~C.~d.~T.,
  {Domingo}, V. \emph{et~al.} 2012 {Supersonic Magnetic Flows in the Quiet Sun
  Observed with SUNRISE/IMaX}.
\newblock In \emph{4th hinode science meeting: Unsolved problems and recent
  insights} (eds L.~{Bellot Rubio}, F.~{Reale} \& M.~{Carlsson}), vol. 455 of
  \emph{Astronomical Society of the Pacific Conference Series}, p. 155.

\bibitem[{{Bradshaw} \& {Cargill}(2006)}]{Bradshaw:2006nx}
{Bradshaw}, S.~J. \& {Cargill}, P.~J. 2006 {Explosive heating of low-density
  coronal plasma}.
\newblock \emph{A\&A}, \textbf{458}, 987--995.
\newblock (\doi{10.1051/0004-6361:20065691})

\bibitem[{{Braginskii}(1965)}]{Braginskii:1965ul}
{Braginskii}, S.~I. 1965 {Transport Processes in a Plasma}.
\newblock \emph{Reviews of Plasma Physics}, \textbf{1}, 205.

\bibitem[{{Brandenburg} \& {Zweibel}(1994)}]{Brandenburg:1994qy}
{Brandenburg}, A. \& {Zweibel}, E.~G. 1994 {The formation of sharp structures
  by ambipolar diffusion}.
\newblock \emph{ApJl}, \textbf{427}, L91--L94.
\newblock (\doi{10.1086/187372})

\bibitem[{{Brandenburg} \& {Zweibel}(1995)}]{Brandenburg:1995jb}
{Brandenburg}, A. \& {Zweibel}, E.~G. 1995 {Effects of Pressure and Resistivity
  on the Ambipolar Diffusion Singularity: Too Little, Too Late}.
\newblock \emph{ApJ}, \textbf{448}, 734.
\newblock (\doi{10.1086/176001})

\bibitem[{{Carlsson} \& {Leenaarts}(2012)}]{Carlsson:2012uq}
{Carlsson}, M. \& {Leenaarts}, J. 2012 {Approximations for radiative cooling
  and heating in the solar chromosphere}.
\newblock \emph{A\&A}, \textbf{539}, A39.
\newblock (\doi{10.1051/0004-6361/201118366})

\bibitem[{{Carlsson} \& {Stein}(1992)}]{Carlsson:1992kl}
{Carlsson}, M. \& {Stein}, R.~F. 1992 {Non-LTE radiating acoustic shocks and CA
  II K2V bright points}.
\newblock \emph{ApJl}, \textbf{397}, L59--L62.
\newblock (\doi{10.1086/186544})

\bibitem[{{Carlsson} \& {Stein}(1995)}]{Carlsson:1995ai}
{Carlsson}, M. \& {Stein}, R.~F. 1995 {Does a nonmagnetic solar chromosphere
  exist?}
\newblock \emph{ApJl}, \textbf{440}, L29--L32.
\newblock (\doi{10.1086/187753})

\bibitem[{{Centeno} \emph{et~al.}(2007){Centeno}, {Socas-Navarro}, {Lites},
  {Kubo}, {Frank}, {Shine}, {Tarbell}, {Title}, {Ichimoto}
  \emph{et~al.}}]{Centeno:2007lr}
{Centeno}, R., {Socas-Navarro}, H., {Lites}, B., {Kubo}, M., {Frank}, Z.,
  {Shine}, R., {Tarbell}, T., {Title}, A., {Ichimoto}, K. \emph{et~al.} 2007
  {Emergence of Small-Scale Magnetic Loops in the Quiet-Sun Internetwork}.
\newblock \emph{ApJl}, \textbf{666}, L137--L140.
\newblock (\doi{10.1086/521726})

\bibitem[{{Cheung} \& {Cameron}(2012)}]{Cheung:2012uq}
{Cheung}, M.~C.~M. \& {Cameron}, R.~H. 2012 {Magnetohydrodynamics of the Weakly
  Ionized Solar Photosphere}.
\newblock \emph{ApJ}, \textbf{750}, 6.
\newblock (\doi{10.1088/0004-637X/750/1/6})

\bibitem[{{Cowling}(1957)}]{cowling1957}
{Cowling}, T.~G. 1957 \emph{Magnetohydrodinamics}.
\newblock Interscience tracts on physics and astronomy.

\bibitem[{{Culhane} \emph{et~al.}(2000){Culhane}, {Korendyke}, {Watanabe} \&
  {Doschek}}]{Culhane:2000lr}
{Culhane}, J.~L., {Korendyke}, C.~M., {Watanabe}, T. \& {Doschek}, G.~A. 2000
  {Extreme-ultraviolet imaging spectrometer designed for the Japanese Solar-B
  satellite}.
\newblock In \emph{Society of photo-optical instrumentation engineers (spie)
  conference series} (ed. {S.~Fineschi, C.~M.~Korendyke, O.~H.~Siegmund, \&
  B.~E.~Woodgate }), vol. 4139 of \emph{Presented at the Society of
  Photo-Optical Instrumentation Engineers (SPIE) Conference}, pp. 294--312.

\bibitem[{{De Moortel} \& {Pascoe}(2012)}]{De-Moortel:2012xe}
{De Moortel}, I. \& {Pascoe}, D.~J. 2012 {The Effects of Line-of-sight
  Integration on Multistrand Coronal Loop Oscillations}.
\newblock \emph{ApJ}, \textbf{746}, 31.
\newblock (\doi{10.1088/0004-637X/746/1/31})

\bibitem[{{De Pontieu}(1999)}]{de-Pontieu:1999kx}
{De Pontieu}, B. 1999 {Numerical simulations of spicules driven by
  weakly-damped Alfv{\'e}n waves. I. WKB approach}.
\newblock \emph{A\&A}, \textbf{347}, 696--710.

\bibitem[{{De Pontieu}(2002)}]{De-Pontieu:2002by}
{De Pontieu}, B. 2002 {High-Resolution Observations of Small-Scale Emerging
  Flux in the Photosphere}.
\newblock \emph{ApJ}, \textbf{569}, 474--486.
\newblock (\doi{10.1086/339231})

\bibitem[{{De Pontieu} \& {Haerendel}(1998)}]{de-Pontieu:1998lr}
{De Pontieu}, B. \& {Haerendel}, G. 1998 {Weakly damped Alfven waves as drivers
  for spicules}.
\newblock \emph{A\&A}, \textbf{338}, 729--736.

\bibitem[{{De Pontieu} \emph{et~al.}(2007{\natexlab{\emph{a}}}){De Pontieu},
  {Hansteen}, {Rouppe van der Voort}, {van Noort} \&
  {Carlsson}}]{De-Pontieu:2007cr}
{De Pontieu}, B., {Hansteen}, V.~H., {Rouppe van der Voort}, L., {van Noort},
  M. \& {Carlsson}, M. 2007{\natexlab{\emph{a}}} {High-Resolution Observations
  and Modeling of Dynamic Fibrils}.
\newblock \emph{ApJ}, \textbf{655}, 624--641.
\newblock (\doi{10.1086/509070})

\bibitem[{{De Pontieu} \emph{et~al.}(2001){De Pontieu}, {Martens} \&
  {Hudson}}]{De-Pontieu:2001fj}
{De Pontieu}, B., {Martens}, P.~C.~H. \& {Hudson}, H.~S. 2001 {Chromospheric
  Damping of Alfv{\'e}n Waves}.
\newblock \emph{ApJ}, \textbf{558}, 859--871.
\newblock (\doi{10.1086/322408})

\bibitem[{{De Pontieu} \emph{et~al.}(2011){De Pontieu}, {McIntosh}, {Carlsson},
  {Hansteen}, {Tarbell}, {Boerner}, {Martinez-Sykora}, {Schrijver} \&
  {Title}}]{De-Pontieu:2011lr}
{De Pontieu}, B., {McIntosh}, S.~W., {Carlsson}, M., {Hansteen}, V.~H.,
  {Tarbell}, T.~D., {Boerner}, P., {Martinez-Sykora}, J., {Schrijver}, C.~J. \&
  {Title}, A.~M. 2011 {The Origins of Hot Plasma in the Solar Corona}.
\newblock \emph{Science}, \textbf{331}, 55--.
\newblock (\doi{10.1126/science.1197738})

\bibitem[{{De Pontieu} \emph{et~al.}(2007{\natexlab{\emph{b}}}){De Pontieu},
  {McIntosh}, {Carlsson}, {Hansteen}, {Tarbell}, {Schrijver}, {Title}, {Shine},
  {Tsuneta} \emph{et~al.}}]{De-Pontieu:2007bd}
{De Pontieu}, B., {McIntosh}, S.~W., {Carlsson}, M., {Hansteen}, V.~H.,
  {Tarbell}, T.~D., {Schrijver}, C.~J., {Title}, A.~M., {Shine}, R.~A.,
  {Tsuneta}, S. \emph{et~al.} 2007{\natexlab{\emph{b}}} {Chromospheric
  Alfv{\'e}nic Waves Strong Enough to Power the Solar Wind}.
\newblock \emph{Science}, \textbf{318}, 1574--.
\newblock (\doi{10.1126/science.1151747})

\bibitem[{{De Pontieu} \emph{et~al.}(2009){De Pontieu}, {McIntosh}, {Hansteen}
  \& {Schrijver}}]{De-Pontieu:2009fk}
{De Pontieu}, B., {McIntosh}, S.~W., {Hansteen}, V.~H. \& {Schrijver}, C.~J.
  2009 {Observing the Roots of Solar Coronal Heating in the Chromosphere}.
\newblock \emph{ApJl}, \textbf{701}, L1--L6.
\newblock (\doi{10.1088/0004-637X/701/1/L1})

\bibitem[{{De Pontieu} \emph{et~al.}(2014){De Pontieu}, {Rouppe van der Voort},
  {McIntosh}, {Pereira}, {Carlsson}, {Hansteen}, {Skogsrud}, {Lemen}, {Title}
  \emph{et~al.}}]{De-Pontieu:2014fv}
{De Pontieu}, B., {Rouppe van der Voort}, L., {McIntosh}, S.~W., {Pereira},
  T.~M.~D., {Carlsson}, M., {Hansteen}, V., {Skogsrud}, H., {Lemen}, J.,
  {Title}, A. \emph{et~al.} 2014 {On the prevalence of small-scale twist in the
  solar chromosphere and transition region}.
\newblock \emph{Science}, \textbf{346}, 1255732.
\newblock (\doi{10.1126/science.1255732})

\bibitem[{{DeForest} \emph{et~al.}(2007){DeForest}, {Hagenaar}, {Lamb},
  {Parnell} \& {Welsch}}]{DeForest:2007ak}
{DeForest}, C.~E., {Hagenaar}, H.~J., {Lamb}, D.~A., {Parnell}, C.~E. \&
  {Welsch}, B.~T. 2007 {Solar Magnetic Tracking. I. Software Comparison and
  Recommended Practices}.
\newblock \emph{ApJ}, \textbf{666}, 576--587.
\newblock (\doi{10.1086/518994})

\bibitem[{{D{\'{\i}}az} \emph{et~al.}(2014){D{\'{\i}}az}, {Khomenko} \&
  {Collados}}]{Diaz:2014dq}
{D{\'{\i}}az}, A.~J., {Khomenko}, E. \& {Collados}, M. 2014 {Rayleigh-Taylor
  instability in partially ionized compressible plasmas: One fluid approach}.
\newblock \emph{A\&A}, \textbf{564}, A97.
\newblock (\doi{10.1051/0004-6361/201322147})

\bibitem[{{Domingo} \emph{et~al.}(1995){Domingo}, {Fleck} \&
  {Poland}}]{Domingo:1995sf}
{Domingo}, V., {Fleck}, B. \& {Poland}, A.~I. 1995 {The SOHO Mission: an
  Overview}.
\newblock \emph{SolPhys}, \textbf{162}, 1--37.
\newblock (\doi{10.1007/BF00733425})

\bibitem[{{Fontenla} \emph{et~al.}(1993){Fontenla}, {Avrett} \&
  {Loeser}}]{Fontenla:1993fj}
{Fontenla}, J.~M., {Avrett}, E.~H. \& {Loeser}, R. 1993 {Energy balance in the
  solar transition region. III - Helium emission in hydrostatic,
  constant-abundance models with diffusion}.
\newblock \emph{ApJ}, \textbf{406}, 319--345.
\newblock (\doi{10.1086/172443})

\bibitem[{{Forteza} \emph{et~al.}(2007){Forteza}, {Oliver}, {Ballester} \&
  {Khodachenko}}]{Forteza:2007zp}
{Forteza}, P., {Oliver}, R., {Ballester}, J.~L. \& {Khodachenko}, M.~L. 2007
  {Damping of oscillations by ion-neutral collisions in a prominence plasma}.
\newblock \emph{A\&A}, \textbf{461}, 731--739.
\newblock (\doi{10.1051/0004-6361:20065900})

\bibitem[{{Galsgaard} \& {Nordlund}(1996)}]{Galsgaard:1996lf}
{Galsgaard}, K. \& {Nordlund}, {\AA}. 1996 {Heating and activity of the solar
  corona 1. Boundary shearing of an initially homogeneous magnetic field}.
\newblock \emph{J.\ Geophys.\ Res.}, \textbf{101}, 13\,445--13\,460.
\newblock (\doi{10.1029/96JA00428})

\bibitem[{{Giannattasio} \emph{et~al.}(2014){Giannattasio}, {Berrilli},
  {Biferale}, {Del Moro}, {Sbragaglia}, {Bellot Rubio}, {Gosic} \& {Orozco
  Suarez}}]{Giannattasio:2014ts}
{Giannattasio}, F., {Berrilli}, F., {Biferale}, L., {Del Moro}, D.,
  {Sbragaglia}, M., {Bellot Rubio}, L., {Gosic}, M. \& {Orozco Suarez}, D. 2014
  {Pair separation of magnetic elements in the quiet Sun}.
\newblock \emph{ArXiv e-prints}.

\bibitem[{{Golding} \emph{et~al.}(2014){Golding}, {Carlsson} \&
  {Leenaarts}}]{Golding:2014fk}
{Golding}, T.~P., {Carlsson}, M. \& {Leenaarts}, J. 2014 {Detailed and
  Simplified Nonequilibrium Helium Ionization in the Solar Atmosphere}.
\newblock \emph{ApJ}, \textbf{784}, 30.
\newblock (\doi{10.1088/0004-637X/784/1/30})

\bibitem[{{G{\"o}m{\"o}ry} \emph{et~al.}(2010){G{\"o}m{\"o}ry}, {Beck},
  {Balthasar}, {Ryb{\'a}k}, {Ku{\v c}era}, {Koza} \&
  {W{\"o}hl}}]{Gomory:2010gb}
{G{\"o}m{\"o}ry}, P., {Beck}, C., {Balthasar}, H., {Ryb{\'a}k}, J., {Ku{\v
  c}era}, A., {Koza}, J. \& {W{\"o}hl}, H. 2010 {Magnetic loop emergence within
  a granule}.
\newblock \emph{A\&A}, \textbf{511}, A14.
\newblock (\doi{10.1051/0004-6361/200912807})

\bibitem[{{Goodman}(2011)}]{Goodman:2011wq}
{Goodman}, M.~L. 2011 {Conditions for Photospherically Driven Alfv{\'e}nic
  Oscillations to Heat the Solar Chromosphere by Pedersen Current Dissipation}.
\newblock \emph{ApJ}, \textbf{735}, 45.
\newblock (\doi{10.1088/0004-637X/735/1/45})

\bibitem[{{Goodman} \& {Judge}(2012)}]{Goodman:2012vn}
{Goodman}, M.~L. \& {Judge}, P.~G. 2012 {Radiating Current Sheets in the Solar
  Chromosphere}.
\newblock \emph{ApJ}, \textbf{751}, 75.
\newblock (\doi{10.1088/0004-637X/751/1/75})

\bibitem[{{Goossens} \emph{et~al.}(2009){Goossens}, {Terradas}, {Andries},
  {Arregui} \& {Ballester}}]{Goossens:2009rw}
{Goossens}, M., {Terradas}, J., {Andries}, J., {Arregui}, I. \& {Ballester},
  J.~L. 2009 {On the nature of kink MHD waves in magnetic flux tubes}.
\newblock \emph{A\&A}, \textbf{503}, 213--223.
\newblock (\doi{10.1051/0004-6361/200912399})

\bibitem[{{Go{\v s}i{\'c}} \emph{et~al.}(2014){Go{\v s}i{\'c}}, {Bellot Rubio},
  {Orozco Su{\'a}rez}, {Katsukawa} \& {Del Toro Iniesta}}]{Gosic:2014ef}
{Go{\v s}i{\'c}}, M., {Bellot Rubio}, L.~R., {Orozco Su{\'a}rez}, D.,
  {Katsukawa}, Y. \& {Del Toro Iniesta}, J.~C. 2014 {The Solar Internetwork. I.
  Contribution to the Network Magnetic Flux}.
\newblock \emph{ArXiv e-prints}.

\bibitem[{{Grossmann-Doerth} \emph{et~al.}(2000){Grossmann-Doerth},
  {Sch{\"u}ssler}, {Sigwarth} \& {Steiner}}]{Grossmann-Doerth:2000wt}
{Grossmann-Doerth}, U., {Sch{\"u}ssler}, M., {Sigwarth}, M. \& {Steiner}, O.
  2000 {Strong Stokes V asymmetries of photospheric spectral lines: What can
  they tell us about the magnetic field structure?}
\newblock \emph{A\&A}, \textbf{357}, 351--358.

\bibitem[{{Gudiksen} \emph{et~al.}(2011){Gudiksen}, {Carlsson}, {Hansteen},
  {Hayek}, {Leenaarts} \& {Mart{\'{\i}}nez-Sykora}}]{Gudiksen:2011qy}
{Gudiksen}, B.~V., {Carlsson}, M., {Hansteen}, V.~H., {Hayek}, W., {Leenaarts},
  J. \& {Mart{\'{\i}}nez-Sykora}, J. 2011 {The stellar atmosphere simulation
  code Bifrost. Code description and validation}.
\newblock \emph{A\&A}, \textbf{531}, A154+.
\newblock (\doi{10.1051/0004-6361/201116520})

\bibitem[{{Gudiksen} \& {Nordlund}(2004)}]{Gudiksen+Nordlund2004}
{Gudiksen}, B.~V. \& {Nordlund}, {\AA}. 2004 {An Ab Initio Approach to the
  Solar Coronal Heating Problem}.
\newblock In \emph{Stars as suns : Activity, evolution and planets} (eds A.~K.
  {Dupree} \& A.~O. {Benz}), vol. 219 of \emph{IAU Symposium}, pp. 488--+.

\bibitem[{{Gudiksen} \&
  {Nordlund}(2005{\natexlab{\emph{a}}})}]{Gudiksen:2005fk}
{Gudiksen}, B.~V. \& {Nordlund}, {\AA}. 2005{\natexlab{\emph{a}}} {An AB Initio
  Approach to Solar Coronal Loops}.
\newblock \emph{ApJ}, \textbf{618}, 1031--1038.
\newblock (\doi{10.1086/426064})

\bibitem[{{Gudiksen} \&
  {Nordlund}(2005{\natexlab{\emph{b}}})}]{Gudiksen:2005lr}
{Gudiksen}, B.~V. \& {Nordlund}, {\AA}. 2005{\natexlab{\emph{b}}} {An Ab Initio
  Approach to the Solar Coronal Heating Problem}.
\newblock \emph{ApJ}, \textbf{618}, 1020--1030.
\newblock (\doi{10.1086/426063})

\bibitem[{{Guglielmino} \emph{et~al.}(2010){Guglielmino}, {Bellot Rubio},
  {Zuccarello}, {Aulanier}, {Vargas Dom{\'{\i}}nguez} \&
  {Kamio}}]{Guglielmino:2010lr}
{Guglielmino}, S.~L., {Bellot Rubio}, L.~R., {Zuccarello}, F., {Aulanier}, G.,
  {Vargas Dom{\'{\i}}nguez}, S. \& {Kamio}, S. 2010 {Multiwavelength
  Observations of Small-scale Reconnection Events Triggered by Magnetic Flux
  Emergence in the Solar Atmosphere}.
\newblock \emph{ApJ}, \textbf{724}, 1083--1098.
\newblock (\doi{10.1088/0004-637X/724/2/1083})

\bibitem[{Hansteen \emph{et~al.}(2014)Hansteen, De~Pontieu, Carlsson, Lemen,
  Title, Boerner, Hurlburt, Tarbell, Wuelser \emph{et~al.}}]{Hansteen17102014}
Hansteen, V., De~Pontieu, B., Carlsson, M., Lemen, J., Title, A., Boerner, P.,
  Hurlburt, N., Tarbell, T.~D., Wuelser, J.~P. \emph{et~al.} 2014 The
  unresolved fine structure resolved: Iris observations of the solar transition
  region.
\newblock \emph{Science}, \textbf{346}(6207).
\newblock (\doi{10.1126/science.1255757})

\bibitem[{{Hansteen} \emph{et~al.}(2007){Hansteen}, {Carlsson} \&
  {Gudiksen}}]{Hansteen:2007dt}
{Hansteen}, V.~H., {Carlsson}, M. \& {Gudiksen}, B. 2007 {3D Numerical Models
  of the Chromosphere, Transition Region, and Corona}.
\newblock In \emph{The physics of chromospheric plasmas} (eds P.~{Heinzel},
  I.~{Dorotovi{\v c}} \& R.~J. {Rutten}), vol. 368 of \emph{Astronomical
  Society of the Pacific Conference Series}, p. 107.

\bibitem[{{Hansteen} \emph{et~al.}(2006){Hansteen}, {De Pontieu}, {Rouppe van
  der Voort}, {van Noort} \& {Carlsson}}]{Hansteen+DePontieu2006}
{Hansteen}, V.~H., {De Pontieu}, B., {Rouppe van der Voort}, L., {van Noort},
  M. \& {Carlsson}, M. 2006 {Dynamic Fibrils Are Driven by Magnetoacoustic
  Shocks}.
\newblock \emph{Apj}, \textbf{647}, L73--L76.
\newblock (\doi{10.1086/507452})

\bibitem[{{Hansteen} \emph{et~al.}(2010){Hansteen}, {Hara}, {De Pontieu} \&
  {Carlsson}}]{Hansteen:2010uq}
{Hansteen}, V.~H., {Hara}, H., {De Pontieu}, B. \& {Carlsson}, M. 2010 {On
  Redshifts and Blueshifts in the Transition Region and Corona}.
\newblock \emph{ApJ}, \textbf{718}, 1070--1078.
\newblock (\doi{10.1088/0004-637X/718/2/1070})

\bibitem[{{Hansteen} \emph{et~al.}(1993){Hansteen}, {Holzer} \&
  {Leer}}]{Hansteen:1993fy}
{Hansteen}, V.~H., {Holzer}, T.~E. \& {Leer}, E. 1993 {Diffusion effects on the
  helium abundance of the solar transition region and corona}.
\newblock \emph{ApJ}, \textbf{402}, 334--343.
\newblock (\doi{10.1086/172137})

\bibitem[{{Hara} \emph{et~al.}(2008){Hara}, {Watanabe}, {Harra}, {Culhane},
  {Young}, {Mariska} \& {Doschek}}]{Hara:2008ad}
{Hara}, H., {Watanabe}, T., {Harra}, L.~K., {Culhane}, J.~L., {Young}, P.~R.,
  {Mariska}, J.~T. \& {Doschek}, G.~A. 2008 {Coronal Plasma Motions near
  Footpoints of Active Region Loops Revealed from Spectroscopic Observations
  with Hinode EIS}.
\newblock \emph{ApJl}, \textbf{678}, L67--L71.
\newblock (\doi{10.1086/588252})

\bibitem[{{Hasan} \& {van Ballegooijen}(2008)}]{Hasan:2008ys}
{Hasan}, S.~S. \& {van Ballegooijen}, A.~A. 2008 {Dynamics of the Solar
  Magnetic Network. II. Heating the Magnetized Chromosphere}.
\newblock \emph{ApJ}, \textbf{680}, 1542--1552.
\newblock (\doi{10.1086/587773})

\bibitem[{{Hayek} \emph{et~al.}(2010){Hayek}, {Asplund}, {Carlsson},
  {Trampedach}, {Collet}, {Gudiksen}, {Hansteen} \& {Leenaarts}}]{Hayek:2010ac}
{Hayek}, W., {Asplund}, M., {Carlsson}, M., {Trampedach}, R., {Collet}, R.,
  {Gudiksen}, B.~V., {Hansteen}, V.~H. \& {Leenaarts}, J. 2010 {Radiative
  transfer with scattering for domain-decomposed 3D MHD simulations of cool
  stellar atmospheres. Numerical methods and application to the quiet,
  non-magnetic, surface of a solar-type star}.
\newblock \emph{A\&A}, \textbf{517}, A49+.
\newblock (\doi{10.1051/0004-6361/201014210})

\bibitem[{{Iida} \emph{et~al.}(2012){Iida}, {Hagenaar} \&
  {Yokoyama}}]{Iida:2012rf}
{Iida}, Y., {Hagenaar}, H.~J. \& {Yokoyama}, T. 2012 {Detection of Flux
  Emergence, Splitting, Merging, and Cancellation of Network Field. I.
  Splitting and Merging}.
\newblock \emph{ApJ}, \textbf{752}, 149.
\newblock (\doi{10.1088/0004-637X/752/2/149})

\bibitem[{{Ishikawa} \emph{et~al.}(2008){Ishikawa}, {Tsuneta}, {Ichimoto},
  {Isobe}, {Katsukawa}, {Lites}, {Nagata}, {Shimizu}, {Shine}
  \emph{et~al.}}]{Ishikawa:2008cq}
{Ishikawa}, R., {Tsuneta}, S., {Ichimoto}, K., {Isobe}, H., {Katsukawa}, Y.,
  {Lites}, B.~W., {Nagata}, S., {Shimizu}, T., {Shine}, R.~A. \emph{et~al.}
  2008 {Transient horizontal magnetic fields in solar plage regions}.
\newblock \emph{A\&A}, \textbf{481}, L25--L28.
\newblock (\doi{10.1051/0004-6361:20079022})

\bibitem[{{Jess} \emph{et~al.}(2009){Jess}, {Mathioudakis}, {Erd{\'e}lyi},
  {Crockett}, {Keenan} \& {Christian}}]{Jess:2009bh}
{Jess}, D.~B., {Mathioudakis}, M., {Erd{\'e}lyi}, R., {Crockett}, P.~J.,
  {Keenan}, F.~P. \& {Christian}, D.~J. 2009 {Alfv{\'e}n Waves in the Lower
  Solar Atmosphere}.
\newblock \emph{Science}, \textbf{323}, 1582--.
\newblock (\doi{10.1126/science.1168680})

\bibitem[{{Joselyn} \emph{et~al.}(1977){Joselyn}, {Munro} \&
  {Holzer}}]{Joselyn:1977th}
{Joselyn}, J.~A., {Munro}, R.~H. \& {Holzer}, T.~E. 1977 {The Validity of
  Ionization Equilibrium in Steady-State Flows.}
\newblock In \emph{Bulletin of the american astronomical society}, vol.~9 of
  \emph{Bulletin of the American Astronomical Society}, p. 650.

\bibitem[{{Kato} \emph{et~al.}(2011){Kato}, {Steiner}, {Steffen} \&
  {Suematsu}}]{Kato:2011dz}
{Kato}, Y., {Steiner}, O., {Steffen}, M. \& {Suematsu}, Y. 2011 {Excitation of
  Slow Modes in Network Magnetic Elements Through Magnetic Pumping}.
\newblock \emph{ApJl}, \textbf{730}, L24.
\newblock (\doi{10.1088/2041-8205/730/2/L24})

\bibitem[{{Khodachenko} \emph{et~al.}(2004){Khodachenko}, {Arber}, {Rucker} \&
  {Hanslmeier}}]{Khodachenko:2004vn}
{Khodachenko}, M.~L., {Arber}, T.~D., {Rucker}, H.~O. \& {Hanslmeier}, A. 2004
  {Collisional and viscous damping of MHD waves in partially ionized plasmas of
  the solar atmosphere}.
\newblock \emph{A\&A}, \textbf{422}, 1073--1084.
\newblock (\doi{10.1051/0004-6361:20034207})

\bibitem[{{Khodachenko} \emph{et~al.}(2006){Khodachenko}, {Rucker}, {Oliver},
  {Arber} \& {Hanslmeier}}]{Khodachenko:2006kx}
{Khodachenko}, M.~L., {Rucker}, H.~O., {Oliver}, R., {Arber}, T.~D. \&
  {Hanslmeier}, A. 2006 {On the mechanisms of MHD wave damping in the partially
  ionized solar plasmas}.
\newblock \emph{Advances in Space Research}, \textbf{37}, 447--455.
\newblock (\doi{10.1016/j.asr.2005.02.025})

\bibitem[{{Khomenko} \& {Collados}(2012)}]{Khomenko:2012bh}
{Khomenko}, E. \& {Collados}, M. 2012 {Heating of the Magnetized Solar
  Chromosphere by Partial Ionization Effects}.
\newblock \emph{ApJ}, \textbf{747}, 87.
\newblock (\doi{10.1088/0004-637X/747/2/87})

\bibitem[{{Khomenko} \emph{et~al.}(2014){Khomenko}, {D{\'{\i}}az}, {de
  Vicente}, {Collados} \& {Luna}}]{Khomenko:2014zr}
{Khomenko}, E., {D{\'{\i}}az}, A., {de Vicente}, A., {Collados}, M. \& {Luna},
  M. 2014 {Rayleigh-Taylor instability in prominences from numerical
  simulations including partial ionization effects}.
\newblock \emph{A\&A}, \textbf{565}, A45.

\bibitem[{{Kjeldseth Moe} \& {Nicolas}(1977)}]{Kjeldseth-Moe:1977lr}
{Kjeldseth Moe}, O. \& {Nicolas}, K.~R. 1977 {Emission measures, electron
  densities, and nonthermal velocities from optically thin UV lines near a
  quiet solar limb}.
\newblock \emph{ApJ}, \textbf{211}, 579--586.
\newblock (\doi{10.1086/154966})

\bibitem[{{Klimchuk} \& {Bradshaw}(2014)}]{Klimchuk:2014fk}
{Klimchuk}, J.~A. \& {Bradshaw}, S.~J. 2014 {Are Chromospheric Nanoflares a
  Primary Source of Coronal Plasma?}
\newblock \emph{ApJ}, \textbf{791}, 60.
\newblock (\doi{10.1088/0004-637X/791/1/60})

\bibitem[{{Kosugi} \emph{et~al.}(2007){Kosugi}, {Matsuzaki}, {Sakao},
  {Shimizu}, {Sone}, {Tachikawa}, {Hashimoto}, {Minesugi}, {Ohnishi}
  \emph{et~al.}}]{2007SoPh..243....3K}
{Kosugi}, T., {Matsuzaki}, K., {Sakao}, T., {Shimizu}, T., {Sone}, Y.,
  {Tachikawa}, S., {Hashimoto}, T., {Minesugi}, K., {Ohnishi}, A. \emph{et~al.}
  2007 {The Hinode (Solar-B) Mission: An Overview}.
\newblock \emph{SolPhys}, \textbf{243}, 3--17.
\newblock (\doi{10.1007/s11207-007-9014-6})

\bibitem[{{Lamb} \emph{et~al.}(2008){Lamb}, {DeForest}, {Hagenaar}, {Parnell}
  \& {Welsch}}]{Lamb:2008hi}
{Lamb}, D.~A., {DeForest}, C.~E., {Hagenaar}, H.~J., {Parnell}, C.~E. \&
  {Welsch}, B.~T. 2008 {Solar Magnetic Tracking. II. The Apparent Unipolar
  Origin of Quiet-Sun Flux}.
\newblock \emph{ApJ}, \textbf{674}, 520--529.
\newblock (\doi{10.1086/524372})

\bibitem[{{Lazarian} \emph{et~al.}(2004){Lazarian}, {Vishniac} \&
  {Cho}}]{Lazarian:2004xi}
{Lazarian}, A., {Vishniac}, E.~T. \& {Cho}, J. 2004 {Magnetic Field Structure
  and Stochastic Reconnection in a Partially Ionized Gas}.
\newblock \emph{ApJ}, \textbf{603}, 180--197.
\newblock (\doi{10.1086/381383})

\bibitem[{Leake \& Arber(2006)}]{Leake2006}
Leake, J.~E. \& Arber, T.~D. 2006 The emergence of magnetic flux through a
  partially ionised solar atmosphere.
\newblock \emph{A\&A}, \textbf{450}, 805--818.

\bibitem[{{Leake} \emph{et~al.}(2005){Leake}, {Arber} \&
  {Khodachenko}}]{Leake:2005rt}
{Leake}, J.~E., {Arber}, T.~D. \& {Khodachenko}, M.~L. 2005 {Collisional
  dissipation of Alfv{\'e}n waves in a partially ionised solar chromosphere}.
\newblock \emph{A\&A}, \textbf{442}, 1091--1098.
\newblock (\doi{10.1051/0004-6361:20053427})

\bibitem[{{Leake} \emph{et~al.}(2013{\natexlab{\emph{a}}}){Leake}, {DeVore},
  {Thayer}, {Burns}, {Crowley}, {Gilbert}, {Huba}, {Judge}, {Krall}
  \emph{et~al.}}]{Leake:2013fk}
{Leake}, J.~E., {DeVore}, C.~R., {Thayer}, J.~P., {Burns}, A.~G., {Crowley},
  G., {Gilbert}, H.~R., {Huba}, J.~D., {Judge}, P., {Krall}, J. \emph{et~al.}
  2013{\natexlab{\emph{a}}} {Ionized Plasma and Neutral Gas Coupling in the
  Sun's Chromosphere and Earth's Ionosphere/Thermosphere}.
\newblock \emph{ArXiv e-prints}.

\bibitem[{{Leake} \& {Linton}(2013)}]{Leake:2013dq}
{Leake}, J.~E. \& {Linton}, M.~G. 2013 {Effect of Ion-Neutral Collisions in
  Simulations of Emerging Active Regions}.
\newblock \emph{ApJ}, \textbf{764}, 54.
\newblock (\doi{10.1088/0004-637X/764/1/54})

\bibitem[{{Leake} \emph{et~al.}(2013{\natexlab{\emph{b}}}){Leake}, {Lukin} \&
  {Linton}}]{Leake:2013qz}
{Leake}, J.~E., {Lukin}, V.~S. \& {Linton}, M.~G. 2013{\natexlab{\emph{b}}}
  {Magnetic reconnection in a weakly ionized plasma}.
\newblock \emph{Physics of Plasmas}, \textbf{20}(6), 061\,202.
\newblock (\doi{10.1063/1.4811140})

\bibitem[{{Leake} \emph{et~al.}(2012){Leake}, {Lukin}, {Linton} \&
  {Meier}}]{Leake:2012pr}
{Leake}, J.~E., {Lukin}, V.~S., {Linton}, M.~G. \& {Meier}, E.~T. 2012
  {Multi-fluid Simulations of Chromospheric Magnetic Reconnection in a Weakly
  Ionized Reacting Plasma}.
\newblock \emph{ApJ}, \textbf{760}, 109.
\newblock (\doi{10.1088/0004-637X/760/2/109})

\bibitem[{{Leenaarts} \emph{et~al.}(2011){Leenaarts}, {Carlsson}, {Hansteen} \&
  {Gudiksen}}]{Leenaarts:2011qy}
{Leenaarts}, J., {Carlsson}, M., {Hansteen}, V. \& {Gudiksen}, B.~V. 2011 {On
  the minimum temperature of the quiet solar chromosphere}.
\newblock \emph{A\&A}, \textbf{530}, A124.
\newblock (\doi{10.1051/0004-6361/201016392})

\bibitem[{{Leenaarts} \emph{et~al.}(2007){Leenaarts}, {Carlsson}, {Hansteen} \&
  {Rutten}}]{Leenaarts:2007sf}
{Leenaarts}, J., {Carlsson}, M., {Hansteen}, V. \& {Rutten}, R.~J. 2007
  {Non-equilibrium hydrogen ionization in 2D simulations of the solar
  atmosphere}.
\newblock \emph{A\&A}, \textbf{473}, 625--632.
\newblock (\doi{10.1051/0004-6361:20078161})

\bibitem[{{Leenaarts} \emph{et~al.}(2012){Leenaarts}, {Carlsson} \& {Rouppe van
  der Voort}}]{Leenaarts:2012cr}
{Leenaarts}, J., {Carlsson}, M. \& {Rouppe van der Voort}, L. 2012 {The
  Formation of the H{$\alpha$} Line in the Solar Chromosphere}.
\newblock \emph{ApJ}, \textbf{749}, 136.
\newblock (\doi{10.1088/0004-637X/749/2/136})

\bibitem[{{Leenaarts} \emph{et~al.}(2013){Leenaarts}, {Pereira}, {Carlsson},
  {Uitenbroek} \& {De Pontieu}}]{Leenaarts:2013ij}
{Leenaarts}, J., {Pereira}, T.~M.~D., {Carlsson}, M., {Uitenbroek}, H. \& {De
  Pontieu}, B. 2013 {The Formation of IRIS Diagnostics. I. A Quintessential
  Model Atom of Mg II and General Formation Properties of the Mg II h and k
  Lines}.
\newblock \emph{ApJ}, \textbf{772}, 89.
\newblock (\doi{10.1088/0004-637X/772/2/89})

\bibitem[{{Lemen} \emph{et~al.}(2012){Lemen}, {Title}, {Akin}, {Boerner},
  {Chou}, {Drake}, {Duncan}, {Edwards}, {Friedlaender}
  \emph{et~al.}}]{Lemen:2012uq}
{Lemen}, J.~R., {Title}, A.~M., {Akin}, D.~J., {Boerner}, P.~F., {Chou}, C.,
  {Drake}, J.~F., {Duncan}, D.~W., {Edwards}, C.~G., {Friedlaender}, F.~M.
  \emph{et~al.} 2012 {The Atmospheric Imaging Assembly (AIA) on the Solar
  Dynamics Observatory (SDO)}.
\newblock \emph{SolPhys}, \textbf{275}, 17--40.
\newblock (\doi{10.1007/s11207-011-9776-8})

\bibitem[{{Lie-Svendsen} \emph{et~al.}(2001){Lie-Svendsen}, {Leer} \&
  {Hansteen}}]{Lie-Svendsen:2001la}
{Lie-Svendsen}, {\O}., {Leer}, E. \& {Hansteen}, V.~H. 2001 {A 16-moment solar
  wind model: From the chromosphere to 1 AU}.
\newblock \emph{J.\ Geophys.\ Res.}, \textbf{106}, 8217--8232.
\newblock (\doi{10.1029/2000JA000409})

\bibitem[{{Lites} \emph{et~al.}(2008){Lites}, {Kubo}, {Socas-Navarro},
  {Berger}, {Frank}, {Shine}, {Tarbell}, {Title}, {Ichimoto}
  \emph{et~al.}}]{Lites:2008ss}
{Lites}, B.~W., {Kubo}, M., {Socas-Navarro}, H., {Berger}, T., {Frank}, Z.,
  {Shine}, R., {Tarbell}, T., {Title}, A., {Ichimoto}, K. \emph{et~al.} 2008
  {The Horizontal Magnetic Flux of the Quiet-Sun Internetwork as Observed with
  the Hinode Spectro-Polarimeter}.
\newblock \emph{ApJ}, \textbf{672}, 1237--1253.
\newblock (\doi{10.1086/522922})

\bibitem[{{Lites} \emph{et~al.}(1996){Lites}, {Leka}, {Skumanich}, {Martinez
  Pillet} \& {Shimizu}}]{Lites:1996kh}
{Lites}, B.~W., {Leka}, K.~D., {Skumanich}, A., {Martinez Pillet}, V. \&
  {Shimizu}, T. 1996 {Small-Scale Horizontal Magnetic Fields in the Solar
  Photosphere}.
\newblock \emph{ApJ}, \textbf{460}, 1019.
\newblock (\doi{10.1086/177028})

\bibitem[{{Lites} \emph{et~al.}(1998){Lites}, {Skumanich} \& {Martinez
  Pillet}}]{Lites:1998cr}
{Lites}, B.~W., {Skumanich}, A. \& {Martinez Pillet}, V. 1998 {Vector magnetic
  fields of emerging solar flux. I. Properties at the site of emergence}.
\newblock \emph{A\&A}, \textbf{333}, 1053--1068.

\bibitem[{{Madsen} \emph{et~al.}(2014){Madsen}, {Dimant}, {Oppenheim} \&
  {Fontenla}}]{Madsen:2014ly}
{Madsen}, C.~A., {Dimant}, Y.~S., {Oppenheim}, M.~M. \& {Fontenla}, J.~M. 2014
  {The Multi-species Farley-Buneman Instability in the Solar Chromosphere}.
\newblock \emph{ApJ}, \textbf{783}, 128.
\newblock (\doi{10.1088/0004-637X/783/2/128})

\bibitem[{{Mart{\'{\i}}nez Gonz{\'a}lez} \& {Bellot
  Rubio}(2009)}]{Martinez-Gonzalez:2009rp}
{Mart{\'{\i}}nez Gonz{\'a}lez}, M.~J. \& {Bellot Rubio}, L.~R. 2009 {Emergence
  of Small-scale Magnetic Loops Through the Quiet Solar Atmosphere}.
\newblock \emph{ApJ}, \textbf{700}, 1391--1403.
\newblock (\doi{10.1088/0004-637X/700/2/1391})

\bibitem[{{Mart{\'{\i}}nez Gonz{\'a}lez} \emph{et~al.}(2012){Mart{\'{\i}}nez
  Gonz{\'a}lez}, {Manso Sainz}, {Asensio Ramos} \&
  {Hijano}}]{Martinez-Gonzalez:2012cl}
{Mart{\'{\i}}nez Gonz{\'a}lez}, M.~J., {Manso Sainz}, R., {Asensio Ramos}, A.
  \& {Hijano}, E. 2012 {Dead Calm Areas in the Very Quiet Sun}.
\newblock \emph{ApJ}, \textbf{755}, 175.
\newblock (\doi{10.1088/0004-637X/755/2/175})

\bibitem[{{Mart{\'{\i}}nez-Sykora} \emph{et~al.}(2012){Mart{\'{\i}}nez-Sykora},
  {De Pontieu} \& {Hansteen}}]{Martinez-Sykora:2012uq}
{Mart{\'{\i}}nez-Sykora}, J., {De Pontieu}, B. \& {Hansteen}, V. 2012
  {Two-dimensional Radiative Magnetohydrodynamic Simulations of the Importance
  of Partial Ionization in the Chromosphere}.
\newblock \emph{ApJ}, \textbf{753}, 161.
\newblock (\doi{10.1088/0004-637X/753/2/161})

\bibitem[{{Mart{\'{\i}}nez-Sykora}
  \emph{et~al.}(2011{\natexlab{\emph{a}}}){Mart{\'{\i}}nez-Sykora}, {De
  Pontieu}, {Hansteen} \& {McIntosh}}]{Martinez-Sykora:2011oq}
{Mart{\'{\i}}nez-Sykora}, J., {De Pontieu}, B., {Hansteen}, V. \& {McIntosh},
  S.~W. 2011{\natexlab{\emph{a}}} {What do Spectral Line Profile Asymmetries
  Tell us About the Solar Atmosphere?}
\newblock \emph{ApJ}, \textbf{732}, 84.
\newblock (\doi{10.1088/0004-637X/732/2/84})

\bibitem[{{Mart{\'{\i}}nez-Sykora} \emph{et~al.}(2013){Mart{\'{\i}}nez-Sykora},
  {De Pontieu}, {Leenaarts}, {Pereira}, {Carlsson}, {Hansteen}, {Stern},
  {Tian}, {McIntosh} \emph{et~al.}}]{Martinez-Sykora:2013ys}
{Mart{\'{\i}}nez-Sykora}, J., {De Pontieu}, B., {Leenaarts}, J., {Pereira},
  T.~M.~D., {Carlsson}, M., {Hansteen}, V., {Stern}, J.~V., {Tian}, H.,
  {McIntosh}, S.~W. \emph{et~al.} 2013 {A Detailed Comparison between the
  Observed and Synthesized Properties of a Simulated Type II Spicule}.
\newblock \emph{ApJ}, \textbf{771}, 66.
\newblock (\doi{10.1088/0004-637X/771/1/66})

\bibitem[{{Mart{\'{\i}}nez-Sykora} \emph{et~al.}(2008){Mart{\'{\i}}nez-Sykora},
  {Hansteen} \& {Carlsson}}]{paper1}
{Mart{\'{\i}}nez-Sykora}, J., {Hansteen}, V. \& {Carlsson}, M. 2008 {Twisted
  Flux Tube Emergence From the Convection Zone to the Corona}.
\newblock \emph{ApJ}, \textbf{679}, 871--888.
\newblock (\doi{10.1086/587028})

\bibitem[{{Mart{\'{\i}}nez-Sykora}
  \emph{et~al.}(2009{\natexlab{\emph{a}}}){Mart{\'{\i}}nez-Sykora}, {Hansteen}
  \& {Carlsson}}]{Martinez-Sykora:2009rw}
{Mart{\'{\i}}nez-Sykora}, J., {Hansteen}, V. \& {Carlsson}, M.
  2009{\natexlab{\emph{a}}} {Twisted Flux Tube Emergence from the Convection
  Zone to the Corona. II. Later States}.
\newblock \emph{ApJ}, \textbf{702}, 129--140.
\newblock (\doi{10.1088/0004-637X/702/1/129})

\bibitem[{{Mart{\'{\i}}nez-Sykora}
  \emph{et~al.}(2009{\natexlab{\emph{b}}}){Mart{\'{\i}}nez-Sykora}, {Hansteen},
  {DePontieu} \& {Carlsson}}]{Martinez-Sykora:2009kl}
{Mart{\'{\i}}nez-Sykora}, J., {Hansteen}, V., {DePontieu}, B. \& {Carlsson}, M.
  2009{\natexlab{\emph{b}}} {Spicule-Like Structures Observed in
  Three-Dimensional Realistic Magnetohydrodynamic Simulations}.
\newblock \emph{ApJ}, \textbf{701}, 1569--1581.
\newblock (\doi{10.1088/0004-637X/701/2/1569})

\bibitem[{{Mart{\'{\i}}nez-Sykora}
  \emph{et~al.}(2011{\natexlab{\emph{b}}}){Mart{\'{\i}}nez-Sykora}, {Hansteen}
  \& {Moreno-Insertis}}]{Martinez-Sykora:2011uq}
{Mart{\'{\i}}nez-Sykora}, J., {Hansteen}, V. \& {Moreno-Insertis}, F.
  2011{\natexlab{\emph{b}}} {On the Origin of the Type II Spicules: Dynamic
  Three-dimensional MHD Simulations}.
\newblock \emph{ApJ}, \textbf{736}, 9--+.
\newblock (\doi{10.1088/0004-637X/736/1/9})

\bibitem[{{McIntosh} \& {De Pontieu}(2009)}]{McIntosh:2009yf}
{McIntosh}, S.~W. \& {De Pontieu}, B. 2009 {Observing Episodic Coronal Heating
  Events Rooted in Chromospheric Activity}.
\newblock \emph{ApJl}, \textbf{706}, L80--L85.
\newblock (\doi{10.1088/0004-637X/706/1/L80})

\bibitem[{{McIntosh} \emph{et~al.}(2011){McIntosh}, {de Pontieu}, {Carlsson},
  {Hansteen}, {Boerner} \& {Goossens}}]{McIntosh:2011fk}
{McIntosh}, S.~W., {de Pontieu}, B., {Carlsson}, M., {Hansteen}, V., {Boerner},
  P. \& {Goossens}, M. 2011 {Alfv{\'e}nic waves with sufficient energy to power
  the quiet solar corona and fast solar wind}.
\newblock \emph{Nat}, \textbf{475}, 477--480.
\newblock (\doi{10.1038/nature10235})

\bibitem[{{Meier}(2011)}]{Meier:2011xd}
{Meier}, E.~T. 2011 {Modeling Plasmas with Strong Anisotropy, Neutral Fluid
  Effects, and Open Boundaries}.
\newblock Ph.D. thesis, University of Washington.

\bibitem[{{Olluri} \emph{et~al.}(2013{\natexlab{\emph{a}}}){Olluri}, {Gudiksen}
  \& {Hansteen}}]{Olluri:2013fu}
{Olluri}, K., {Gudiksen}, B.~V. \& {Hansteen}, V.~H. 2013{\natexlab{\emph{a}}}
  {Non-equilibrium Ionization Effects on the Density Line Ratio Diagnostics of
  O IV}.
\newblock \emph{ApJ}, \textbf{767}, 43.
\newblock (\doi{10.1088/0004-637X/767/1/43})

\bibitem[{{Olluri} \emph{et~al.}(2013{\natexlab{\emph{b}}}){Olluri}, {Gudiksen}
  \& {Hansteen}}]{Olluri:2013uq}
{Olluri}, K., {Gudiksen}, B.~V. \& {Hansteen}, V.~H. 2013{\natexlab{\emph{b}}}
  {Non-equilibrium Ionization in the Bifrost Stellar Atmosphere Code}.
\newblock \emph{AJ}, \textbf{145}, 72.
\newblock (\doi{10.1088/0004-6256/145/3/72})

\bibitem[{{Ortiz} \emph{et~al.}(2014){Ortiz}, {Bellot Rubio}, {Hansteen}, {de
  la Cruz Rodr{\'{\i}}guez} \& {Rouppe van der Voort}}]{Ortiz:2014wj}
{Ortiz}, A., {Bellot Rubio}, L.~R., {Hansteen}, V.~H., {de la Cruz
  Rodr{\'{\i}}guez}, J. \& {Rouppe van der Voort}, L. 2014 {Emergence of
  Granular-sized Magnetic Bubbles through the Solar Atmosphere. I.
  Spectropolarimetric Observations and Simulations}.
\newblock \emph{ApJ}, \textbf{781}, 126.
\newblock (\doi{10.1088/0004-637X/781/2/126})

\bibitem[{{Osterbrock}(1961)}]{Osterbrock:1961fk}
{Osterbrock}, D.~E. 1961 {The Heating of the Solar Chromosphere, Plages, and
  Corona by Magnetohydrodynamic Waves.}
\newblock \emph{ApJ}, \textbf{134}, 347--+.
\newblock (\doi{10.1086/147165})

\bibitem[{{Pandey} \& {Wardle}(2008)}]{Pandey:2008qy}
{Pandey}, B.~P. \& {Wardle}, M. 2008 {Hall magnetohydrodynamics of partially
  ionized plasmas}.
\newblock \emph{MNRAS}, \textbf{385}, 2269--2278.
\newblock (\doi{10.1111/j.1365-2966.2008.12998.x})

\bibitem[{{Parker}(1983)}]{Parker:1983fj}
{Parker}, E.~N. 1983 {Magnetic Neutral Sheets in Evolving Fields - Part Two -
  Formation of the Solar Corona}.
\newblock \emph{ApJ}, \textbf{264}, 642.
\newblock (\doi{10.1086/160637})

\bibitem[{{Parker}(2007)}]{Parker:2007lr}
{Parker}, E.~N. 2007 \emph{{Conversations on Electric and Magnetic Fields in
  the Cosmos}}.
\newblock Princeton University Press.

\bibitem[{{Penn} \emph{et~al.}(2011){Penn}, {Schad} \& {Cox}}]{Penn:2011fk}
{Penn}, M.~J., {Schad}, T. \& {Cox}, E. 2011 {Probing the Solar Atmosphere
  Using Oscillations of Infrared CO Spectral Lines}.
\newblock \emph{ApJ}, \textbf{734}, 47.
\newblock (\doi{10.1088/0004-637X/734/1/47})

\bibitem[{{Pesnell} \emph{et~al.}(2012){Pesnell}, {Thompson} \&
  {Chamberlin}}]{Pesnell:2012nr}
{Pesnell}, W.~D., {Thompson}, B.~J. \& {Chamberlin}, P.~C. 2012 {The Solar
  Dynamics Observatory (SDO)}.
\newblock \emph{SolPhys}, \textbf{275}, 3--15.
\newblock (\doi{10.1007/s11207-011-9841-3})

\bibitem[{{Peter}(2001)}]{Peter:2001qy}
{Peter}, H. 2001 {On the nature of the transition region from the chromosphere
  to the corona of the Sun}.
\newblock \emph{A\&A}, \textbf{374}, 1108--1120.
\newblock (\doi{10.1051/0004-6361:20010697})

\bibitem[{{Peter}(2010)}]{Peter:2010fk}
{Peter}, H. 2010 {Asymmetries of solar coronal extreme ultraviolet emission
  lines}.
\newblock \emph{A\&A}, \textbf{521}, A51+.
\newblock (\doi{10.1051/0004-6361/201014433})

\bibitem[{{Peter} \emph{et~al.}(2006){Peter}, {Gudiksen} \&
  {Nordlund}}]{Peter:2006zk}
{Peter}, H., {Gudiksen}, B.~V. \& {Nordlund}, {\AA}. 2006 {Forward Modeling of
  the Corona of the Sun and Solar-like Stars: From a Three-dimensional
  Magnetohydrodynamic Model to Synthetic Extreme-Ultraviolet Spectra}.
\newblock \emph{ApJ}, \textbf{638}, 1086--1100.
\newblock (\doi{10.1086/499117})

\bibitem[{{Roberts} \& {Webb}(1978)}]{Roberts1978}
{Roberts}, B. \& {Webb}, A.~R. 1978 Vertical motions in an intense magnetic
  flux tube.
\newblock \emph{Sol. Phys.}, \textbf{56}, 5--35.

\bibitem[{{Sainz Dalda} \emph{et~al.}(2012){Sainz Dalda},
  {Mart{\'{\i}}nez-Sykora}, {Bellot Rubio} \& {Title}}]{Sainz-Dalda:2012qf}
{Sainz Dalda}, A., {Mart{\'{\i}}nez-Sykora}, J., {Bellot Rubio}, L. \& {Title},
  A. 2012 {Study of Single-lobed Circular Polarization Profiles in the Quiet
  Sun}.
\newblock \emph{ApJ}, \textbf{748}, 38.
\newblock (\doi{10.1088/0004-637X/748/1/38})

\bibitem[{{Sakai} \& {Smith}(2009)}]{Sakai:2009le}
{Sakai}, J.~I. \& {Smith}, P.~D. 2009 {Two-Fluid Simulations of Coalescing
  Penumbra Filaments Driven by Neutral-Hydrogen Flows}.
\newblock \emph{ApJl}, \textbf{691}, L45--L48.
\newblock (\doi{10.1088/0004-637X/691/1/L45})

\bibitem[{{Schrijver} \emph{et~al.}(1997){Schrijver}, {Title}, {van
  Ballegooijen}, {Hagenaar} \& {Shine}}]{Schrijver:1997xu}
{Schrijver}, C.~J., {Title}, A.~M., {van Ballegooijen}, A.~A., {Hagenaar},
  H.~J. \& {Shine}, R.~A. 1997 {Sustaining the Quiet Photospheric Network: The
  Balance of Flux Emergence, Fragmentation, Merging, and Cancellation}.
\newblock \emph{ApJ}, \textbf{487}, 424--436.

\bibitem[{{Skartlien} \emph{et~al.}(2000){Skartlien}, {Stein} \&
  {Nordlund}}]{Skartlien:2000lr}
{Skartlien}, R., {Stein}, R.~F. \& {Nordlund}, {\AA}. 2000 {Excitation of
  Chromospheric Wave Transients by Collapsing Granules}.
\newblock \emph{ApJ}, \textbf{541}, 468--488.
\newblock (\doi{10.1086/309414})

\bibitem[{{Smith} \& {Sakai}(2008)}]{Smith:2008oa}
{Smith}, P.~D. \& {Sakai}, J.~I. 2008 {Chromospheric magnetic reconnection:
  two-fluid simulations of coalescing current loops}.
\newblock \emph{A\&A}, \textbf{486}, 569--575.
\newblock (\doi{10.1051/0004-6361:200809624})

\bibitem[{{Solanki} \emph{et~al.}(2010){Solanki}, {Barthol}, {Danilovic},
  {Feller}, {Gandorfer}, {Hirzberger}, {Riethm{\"u}ller}, {Sch{\"u}ssler},
  {Bonet} \emph{et~al.}}]{Solanki:2010pt}
{Solanki}, S.~K., {Barthol}, P., {Danilovic}, S., {Feller}, A., {Gandorfer},
  A., {Hirzberger}, J., {Riethm{\"u}ller}, T.~L., {Sch{\"u}ssler}, M., {Bonet},
  J.~A. \emph{et~al.} 2010 {SUNRISE: Instrument, Mission, Data, and First
  Results}.
\newblock \emph{ApJl}, \textbf{723}, L127--L133.
\newblock (\doi{10.1088/2041-8205/723/2/L127})

\bibitem[{{Soler} \emph{et~al.}(2012){Soler}, {Andries} \&
  {Goossens}}]{Soler:2012hb}
{Soler}, R., {Andries}, J. \& {Goossens}, M. 2012 {Resonant Alfv{\'e}n waves in
  partially ionized plasmas of the solar atmosphere}.
\newblock \emph{A\&A}, \textbf{537}, A84.
\newblock (\doi{10.1051/0004-6361/201118235})

\bibitem[{{Soler} \emph{et~al.}(2009){Soler}, {Oliver} \&
  {Ballester}}]{Soler:2009hw}
{Soler}, R., {Oliver}, R. \& {Ballester}, J.~L. 2009 {Magnetohydrodynamic Waves
  in a Partially Ionized Filament Thread}.
\newblock \emph{ApJ}, \textbf{699}, 1553--1562.
\newblock (\doi{10.1088/0004-637X/699/2/1553})

\bibitem[{{Song} \& {Vasyli{\=u}nas}(2011)}]{Song:2011ly}
{Song}, P. \& {Vasyli{\=u}nas}, V.~M. 2011 {Heating of the solar atmosphere by
  strong damping of Alfv{\'e}n waves}.
\newblock \emph{Journal of Geophysical Research (Space Physics)}, \textbf{116},
  A09104.
\newblock (\doi{10.1029/2011JA016679})

\bibitem[{{Spruit}(1981)}]{spruit1981}
{Spruit}, H.~C. 1981 Motion of magnetic flux tubes in the solar convection zone
  and chromosphere.
\newblock \emph{A\&A}, \textbf{98}, 155--160.

\bibitem[{{Stein} \& {Nordlund}(2006)}]{Stein:2006qy}
{Stein}, R.~F. \& {Nordlund}, {\AA}. 2006 {Solar Small-Scale
  Magnetoconvection}.
\newblock \emph{ApJ}, \textbf{642}, 1246--1255.
\newblock (\doi{10.1086/501445})

\bibitem[{{Stenflo}(2013)}]{Stenflo:2013fc}
{Stenflo}, J.~O. 2013 {Horizontal or vertical magnetic fields on the quiet Sun.
  Angular distributions and their height variations}.
\newblock \emph{A\&A}, \textbf{555}, A132.
\newblock (\doi{10.1051/0004-6361/201321608})

\bibitem[{Testa \emph{et~al.}(2014)Testa, De~Pontieu, Allred, Carlsson, Reale,
  Daw, Hansteen, Martinez-Sykora, Liu \emph{et~al.}}]{Testa17102014}
Testa, P., De~Pontieu, B., Allred, J., Carlsson, M., Reale, F., Daw, A.,
  Hansteen, V., Martinez-Sykora, J., Liu, W. \emph{et~al.} 2014 Evidence of
  nonthermal particles in coronal loops heated impulsively by nanoflares.
\newblock \emph{Science}, \textbf{346}(6207).
\newblock (\doi{10.1126/science.1255724})

\bibitem[{{Tian} \emph{et~al.}(2011){Tian}, {McIntosh}, {De Pontieu},
  {Mart{\'{\i}}nez-Sykora}, {Sechler} \& {Wang}}]{Tian:2011dq}
{Tian}, H., {McIntosh}, S.~W., {De Pontieu}, B., {Mart{\'{\i}}nez-Sykora}, J.,
  {Sechler}, M. \& {Wang}, X. 2011 {Two Components of the Solar Coronal
  Emission Revealed by Extreme-ultraviolet Spectroscopic Observations}.
\newblock \emph{ApJ}, \textbf{738}, 18.
\newblock (\doi{10.1088/0004-637X/738/1/18})

\bibitem[{{Tsuneta} \emph{et~al.}(2008){Tsuneta}, {Ichimoto}, {Katsukawa},
  {Nagata}, {Otsubo}, {Shimizu}, {Suematsu}, {Nakagiri}, {Noguchi}
  \emph{et~al.}}]{Tsuneta:2008kc}
{Tsuneta}, S., {Ichimoto}, K., {Katsukawa}, Y., {Nagata}, S., {Otsubo}, M.,
  {Shimizu}, T., {Suematsu}, Y., {Nakagiri}, M., {Noguchi}, M. \emph{et~al.}
  2008 {The Solar Optical Telescope for the Hinode Mission: An Overview}.
\newblock \emph{SolPhys}, \textbf{249}, 167--196.
\newblock (\doi{10.1007/s11207-008-9174-z})

\bibitem[{{van Ballegooijen} \emph{et~al.}(2011){van Ballegooijen},
  {Asgari-Targhi}, {Cranmer} \& {DeLuca}}]{van-Ballegooijen:2011fp}
{van Ballegooijen}, A.~A., {Asgari-Targhi}, M., {Cranmer}, S.~R. \& {DeLuca},
  E.~E. 2011 {Heating of the Solar Chromosphere and Corona by Alfv{\'e}n Wave
  Turbulence}.
\newblock \emph{ApJ}, \textbf{736}, 3.
\newblock (\doi{10.1088/0004-637X/736/1/3})

\bibitem[{{Vishniac} \& {Lazarian}(1999)}]{Vishniac:1999zk}
{Vishniac}, E.~T. \& {Lazarian}, A. 1999 {Reconnection in the Interstellar
  Medium}.
\newblock \emph{ApJ}, \textbf{511}, 193--203.
\newblock (\doi{10.1086/306643})

\bibitem[{{Viticchi{\'e}}(2012)}]{Viticchie:2012ez}
{Viticchi{\'e}}, B. 2012 {On the Polarimetric Signature of Emerging Magnetic
  Loops in the Quiet Sun}.
\newblock \emph{ApJl}, \textbf{747}, L36.
\newblock (\doi{10.1088/2041-8205/747/2/L36})

\bibitem[{{Viticchi{\'e}} \emph{et~al.}(2011){Viticchi{\'e}}, {S{\'a}nchez
  Almeida}, {Del Moro} \& {Berrilli}}]{Viticchie:2011si}
{Viticchi{\'e}}, B., {S{\'a}nchez Almeida}, J., {Del Moro}, D. \& {Berrilli},
  F. 2011 {Interpretation of HINODE SOT/SP asymmetric Stokes profiles observed
  in the quiet Sun network and internetwork}.
\newblock \emph{A\&A}, \textbf{526}, A60.
\newblock (\doi{10.1051/0004-6361/201015391})

\bibitem[{{von Steiger} \& {Geiss}(1989)}]{von-Steiger:1989uq}
{von Steiger}, R. \& {Geiss}, J. 1989 {Supply of fractionated gases to the
  corona}.
\newblock \emph{A\&A}, \textbf{225}, 222--238.

\bibitem[{{Vranjes} \& {Kono}(2014)}]{Vranjes:2014fk}
{Vranjes}, J. \& {Kono}, M. 2014 {On the Alfv{\'e}n wave cut-off in partly
  ionized collisional plasmas}.
\newblock \emph{Physics of Plasmas}, \textbf{21}(1), 012110.
\newblock (\doi{10.1063/1.4862781})

\bibitem[{{Vranjes} \& {Krstic}(2013)}]{Vranjes:2013ve}
{Vranjes}, J. \& {Krstic}, P.~S. 2013 {Collisions, magnetization, and transport
  coefficients in the lower solar atmosphere}.
\newblock \emph{A\&A}, \textbf{554}, A22.
\newblock (\doi{10.1051/0004-6361/201220738})

\bibitem[{{Wilhelm} \emph{et~al.}(1995){Wilhelm}, {Curdt}, {Marsch},
  {Sch{\"u}hle}, {Lemaire}, {Gabriel}, {Vial}, {Grewing}, {Huber}
  \emph{et~al.}}]{Wilhelm:1995fk}
{Wilhelm}, K., {Curdt}, W., {Marsch}, E., {Sch{\"u}hle}, U., {Lemaire}, P.,
  {Gabriel}, A., {Vial}, J., {Grewing}, M., {Huber}, M.~C.~E. \emph{et~al.}
  1995 {SUMER - Solar Ultraviolet Measurements of Emitted Radiation}.
\newblock \emph{SolPhys}, \textbf{162}, 189--231.
\newblock (\doi{10.1007/BF00733430})

\bibitem[{{Zaqarashvili} \emph{et~al.}(2012){Zaqarashvili}, {Carbonell},
  {Ballester} \& {Khodachenko}}]{Zaqarashvili:2012iw}
{Zaqarashvili}, T.~V., {Carbonell}, M., {Ballester}, J.~L. \& {Khodachenko},
  M.~L. 2012 {Cut-off wavenumber of Alfv{\'e}n waves in partially ionized
  plasmas of the solar atmosphere}.
\newblock \emph{A\&A}, \textbf{544}, A143.
\newblock (\doi{10.1051/0004-6361/201219763})

\bibitem[{{Zhou} \emph{et~al.}(2013){Zhou}, {Wang} \& {Jin}}]{Zhou:2013bl}
{Zhou}, G., {Wang}, J. \& {Jin}, C. 2013 {Solar Intranetwork Magnetic Elements:
  Flux Distributions}.
\newblock \emph{SolPhys}, \textbf{283}, 273--282.
\newblock (\doi{10.1007/s11207-013-0229-4})

\bibitem[{{Zhou} \emph{et~al.}(2010){Zhou}, {Wang} \& {Jin}}]{Zhou:2010vh}
{Zhou}, G.~P., {Wang}, J.~X. \& {Jin}, C.~L. 2010 {Solar Intranetwork Magnetic
  Elements: Evolution and Lifetime}.
\newblock \emph{SolPhys}, \textbf{267}, 63--73.
\newblock (\doi{10.1007/s11207-010-9641-1})

\end{thebibliography}

\end{document}